\documentclass[letter,11pt]{article}
\usepackage[textwidth=15.2cm,textheight=22cm]{geometry}
\usepackage{amsmath,amsfonts,amssymb,color,psfrag,graphicx,fancyhdr,array}
\usepackage[latin1]{inputenc}

\usepackage{bbm}
\usepackage{amssymb}
\usepackage{amsmath}
\usepackage{amscd}
\usepackage{pst-all}
\usepackage{pst-node}

\usepackage{ifthen}

\def\a{\alpha}
\def\b{\beta}

\def\d{\delta}
\def\e{\eta}
\def\eps{\epsilon}
\def\ve{\varepsilon}

\def\k{\kappa}

\def\la{\lambda}

\newcommand{\ad}{{\dot{\alpha}}}
\newcommand{\bd}{{\dot{\beta}}}

\newcommand{\lb}{{\bar{\lambda}}}

\newcommand{\ba}{{\mathbf{a}}}

\newcommand{\ZC}{\mathbbm C}

\newcommand{\CN}{\mathcal{N}}      
\newcommand{\CO}{\mathcal{O}}      
\newcommand{\CS}{\mathcal{S}}

\def\_{\;\;}
\def\^{\wedge}

\def\pd{\mbox{$\partial$}}

\def\eqn#1{eq.~(\ref{#1})}
\def\Eqn#1{Equation~(\ref{#1})}
\def\eqns#1#2{eqs.~(\ref{#1}) and~(\ref{#2})}

\def\sfrac#1#2{{\textstyle\frac{#1}{#2}}}
\def\>{\rangle}
\def\<{\langle}
\def\+{\dagger}
\def\={\ =\ }

\def\and{\qquad\textrm{and}\qquad}

\def\ki#1{[\![#1]\!]}
\newcommand{\Nf}{{\ensuremath{\mathcal N{=}4}\ }}
\newcommand{\Ne}{{\ensuremath{\mathcal N{=}8}\ }}

\newcommand{\nnl}{\nonumber\\}

\def\be{\begin{equation}}
\def\ee{\end{equation}}

\def\ba{\begin{array}}
\def\ea{\end{array}}

\def\blfootnote{\xdef\@thefnmark{}\@footnotetext}

\numberwithin{equation}{section}

\begin{document}


\begin{titlepage}
\hfill{}

\begin{flushright}
SU-ITP-11/05\\
\today
\end{flushright}

\vskip 2cm

\vspace{24pt}

\begin{center}
{ \LARGE {\bf    From lightcone actions to maximally  }}\\[4pt]
{ \LARGE {\bf  supersymmetric amplitudes}}

\vspace{24pt}

{\large  {\bf  Johannes Broedel${}^\dag$ and   Renata Kallosh${}^\ddag$}}

\vspace{15pt}

{${}^\dag$,${}^\ddag$\,Department of Physics, Stanford University, Stanford, CA 94305}
\end{center}
\vspace{14pt}
\begin{abstract}

In this article actions for \Nf SYM and \Ne supergravity are formulated in terms of a chiral superfield, which contains only the physical degrees of freedom of either theory. In these new actions, which originate from the lightcone superspace,  the supergravity cubic vertex is the square of the gauge theory one (omitting the color structures). Amplitude calculations using the corresponding Feynman supergraph rules are tedious, but can be simplified by choosing a preferred superframe. Recursive calculations of all MHV amplitudes in \Nf SYM and the four-point \Ne supergravity amplitude are shown to agree with the known results and connections to the BCFW recursion relations are pointed out. Finally, the new path integrals are discussed in the context of the double-copy property relating \Nf SYM theory to \Ne supergravity. 

\end{abstract}
\vfill
\psset{xunit=4pt,yunit=4pt,linewidth=0.7pt}
\begin{pspicture}(45,1)
\psline{-}(0,0)(43.5,0)
\end{pspicture}\\
{\footnotesize{\indent E-mail: ${}^\dag$jbroedel@stanford.edu, ${}^\ddag$kallosh@stanford.edu}}
\end{titlepage}


\newpage
\pagestyle{empty}
\tableofcontents


\newpage
\pagestyle{plain}
\setcounter{page}{1}
\section{Introduction}
Remarkable progress has been made for maximally supersymmetric field theories during the last few years \cite{Bern:2007hh,Bern:2009kd}. Many new findings about \Nf super Yang-Mills (SYM) theory trace back to the application of twistor methods and the exploration of the dual conformal symmetry combined with unitarity \cite{Bern:2007ct,Drummond:2008vq,Drummond:2008bq,ArkaniHamed:2010kv}. For \Ne supergravity a natural and equally geometrical approach along the lines of Grassmannian approach to \Nf SYM is still missing.

Most of the recent findings mentioned above have been performed without making use of the path integral formulation of the theories and rather relied on the unitarity and the symmetries of the S-matrix. In terms of Feynman-graph calculations the manifestly Lorentz covariant spacetime formulations are algebraically involved.
One of the difficulties of the Lagrangian approach is the lack of a formulation, which is manifestly invariant under all supersymmetries and which simultaneously exhibits manifest Lorentz invariance. However, paying the price of non-manifest Lorentz-symmetry, one can make eight/sixteen supersymmetries of \Nf SYM/\Ne supergravity manifest in the lightcone formalism \cite{Brink:1982pd,Mandelstam:1982cb,Bengtsson:1983pg,Ananth:2005vg,Brink:2008qc}.

In this article the lightcone formalism will be employed: starting from the unitary and manifestly supersymmetric, but not Lorentz covariant lightcone path integrals for \Nf SYM and \Ne supergravity, a change of variable and a Fourier transformation will be performed\footnote{For \Nf SYM a similar procedure has been performed by Belitsky, Derkachov, Korchemsky and Manashov (BDKM) in \cite{Belitsky:2004sc}.  The results are algebraically equivalent, which will be shown in the appendix explicitly.} clarifying and developing the program initiated in \cite{Kallosh:2009db,Fu:2010qi}. The resulting actions are thereby formulated in terms of just one chiral scalar superfield, which corresponds to the CPT-selfconjugate supermultiplet of physical states. The actions contain Lorentz-covariant cubic interactions corresponding to the seeds of a recursion relation for both theories. In addition, there are quartic contact terms in \Nf SYM and, probably, all higher-point contact terms in \Ne supergravity.

More explicitely, the off-shell three-vertices take the form of the three-point MHV and $\overline{\text{MHV}}$ interactions, whose supersymmetric forms were proposed in \cite{ArkaniHamed:2008gz,Brandhuber:2008pf,Drummond:2008cr}. Given their formulation in terms of usual spinor-helicity brackets, the external legs have to be on-shell. However, for employing the vertices in a path integral, one needs the internal fields -- and thus those attached to the vertices -- to be off-shell. We will address this issue using an lightcone off-shell extension of the spinor helicity formalism.

Performing amplitude calculations based on those path integrals, one still faces the algebraic difficulties of a non-manifest Lorentz-symmetric formulation. However, this apparent disadvantage will be used here to deduce a complex deformation of all external super-momenta, which greatly simplifies the calculation. The implications are similar to the situation after performing  the Britto-Cachazo-Feng-Witten (BCFW)-shift \cite{Britto:2004ap,Britto:2005fq}, a supersymmetric version of which was
proposed in \cite{ArkaniHamed:2008gz,Brandhuber:2008pf,Drummond:2008cr}: once the deformation is applied, many Feynman supergraphs do not contribute to the calculation, because their super-momentum conservation can not be satisfied in the deformed scenario\footnote{In \cite{Chalmers:1998jb} another convenient approach connecting the lightcone formalism to spinor helicity has been discussed.}. This will be shown explicitly in the example of the four-point function in \Nf SYM (proving a qualitative earlier result in \cite{Fu:2010qi}). In particular, one finds that the 4-point contact term does not contribute to the amplitude when a specific choice of a superframe is made. 

Generalizations to more legs and the application to \Ne supergravity turn out to be straightforward. In this sense the path integrals for maximally supersymmetric theories in its new form are suggestive of the existence of on-shell recursion relations.

The article is organized as follows: section \ref{scn:LCformalism} introduces the lightcone formalism, treats the off-shell extension of the spinor-helicity formalism and presents the new forms of the path integrals for \Nf SYM and \Ne supergravity. In section \ref{scn:LCcov} a few subtleties related to the choice of lightcone coordinates, spacetime signatures and complexification of external momenta are discussed. Section \ref{scn:amplitudes} is dedicatted to the calculation of several amplitudes in either theories. The choice of a preferred superframe in order to make the calculation feasible is elucidated and compared to the BCFW deformation. Finally, in the last section ``gravity as a double copy of gauge theory'' is discussed in the context of the new the path integrals for \Nf SYM and \Ne supergravity.


\section{From the lightcone superspace formalism to new path integrals}\label{scn:LCformalism}

In order to transform the known forms of lightcone path integrals into a novel form, one has to employ a procedure, which has been proposed in \cite{Kallosh:2009db,Fu:2010qi}.  In this section the conventions for the lightcone formalism are given and the procedure is reviewed briefly. While the previous result for \Nf SYM theory will be stated, for \Ne supergravity, one example of a conversion will be performed explicitely before stating the new form of the \Ne supergravity action.


\subsection{Conventions}
Ligth-cone corrdinates and corresponding derivatives are defined as
\begin{eqnarray}
x_{\pm}=(x_{0}\pm x_{3})/\sqrt{2}\quad\text{and}\quad x_{\bot}=(x_{1}+ix_{2})/\sqrt{2},\,\,\bar{x}_{\bot}=(x_{1}-ix_{2})/\sqrt{2}\nnl
\pd_{\pm}=\frac{\pd}{\pd x_{\mp}}=\frac{1}{\sqrt{2}}(\partial_{x_{0}}-\partial_{x_{3}}),\,\,\partial_{\bot}=\frac{\partial}{\partial\bar{x}_{\bot}}=\frac{1}{\sqrt{2}}(\partial_{x_{1}}+i\partial_{x_{2}}),\,\,\bar{\partial}_{\bot}=\frac{\partial}{\partial x_{\bot}}.
\end{eqnarray}
With the flat metric being off-diagonal in lightcone coordinates, the scalar product
of two 4-vectors $p$ and $q$ reads
\begin{equation}\label{eqn:scalarproduct}
p\cdot q=p_{+}q_{-}+p_{-}q_{+}-p_{\bot}\bar{q}_{\bot}-\bar{p}_{\bot}q_{\bot}\quad\text{and thus}\quad p^2=2(p_+p_--p_\bot\bar p_\bot)\,.
\end{equation}
A general non-lightlike vector can be decomposed into spinors as
\begin{equation}\label{eqn:vectordecomposition}
p_{\a\ad}=\la_\a\bar\la_\ad+\xi_\a\bar\xi_\ad\,,
\end{equation}
where $\la_\a\bar\la_\ad$ is the on-shell part and $\xi_\a\bar\xi_\ad$ is the off-shell correction. A convenient choice for the holomorphic and anti-holomorphic parts of those spinors is
\begin{equation}\label{eqn:spinorchoice}
\la_\a = \frac{2^{1/4}}{\sqrt{p_+}} \left(
\begin{array}{c}
-p_{\bot} \\
p_+ \\
\end{array}
\right)\, , \qquad
\bar\la_{\ad} = 2^{1/4}\sqrt{p_+}\left(
\begin{array}{c}
-\frac{\bar p_{\bot}}{p_+} \\
1\\
\end{array} \right)\,
\end{equation}
\begin{equation}
\xi_\a = \frac{1}{2^{1/4}\sqrt{p_+}} \left(
\begin{array}{c}
\sqrt{p^2}\\
0 \\
\end{array}
\right)\, , \qquad
\bar\xi_{\ad} = \frac{\sqrt{p_+}}{2^{1/4}p_+}\left(
\begin{array}{c}
\sqrt{p^2}\\
0\\
\end{array} \right)\,.
\end{equation}
For negative $p_{+}$ its square root is defined as $\sqrt{p_{+}}\equiv sgn(p_{+})\,\left|p_{+}\right|^{1/2}$. Accordingly,
$\bar{\lambda}$ picks up a minus sign when $p$ is reversed while $\lambda$ remains unchanged:
\begin{equation}\label{eqn:lachange}
\lambda_{\alpha}(-p)=\lambda_{\alpha}(p),\,\,\,\bar{\lambda}_{\dot{\alpha}}(-p)=-\bar{\lambda}_{\dot{\alpha}}(p).
\end{equation}
Employing the above definitions one finds
\begin{equation}
\la_\a\bar\la_\ad  =  \sqrt{2}\left( \begin{array}{cc} {p_\bot \bar p_\bot  \over p_+}  &  -p_\bot\\
-\bar p_\bot & p_+\\ \end{array} \right)\, ,
\qquad
\xi_\a\bar\xi_\ad  =  \sqrt{2}\left( \begin{array}{cc} \frac{p^2}{2p_+}  &  0\\
 0 & 0\\ \end{array} \right)\,
\end{equation}
and thus
\begin{equation}\label{eqn:momentummatrix}
\xi_\a\bar\xi_\ad+ \la_\a\bar\la_\ad  =  \sqrt{2}\left( \begin{array}{cc}  p_- &  -p_\bot \\
-\bar p_\bot  & p_+\\ \end{array} \right).
\end{equation}
For an on-shell vector ($p^2=0$), the spinors $\xi$ and $\bar\xi$ will vanish, thus leading to the usual on-shell decomposition into two spinors $\la$ and $\bar\la$. \footnote{This convention is the same as Dixon's choice in \cite{Dixon:1996wi}.}


\subsection{Spinor helicity formalism and the lightcone off-shell extension}

Although angular and square spinor brackets are usually defined for massless objects, it is possible to extend the definition to include massive particles as well. This traces back to the fact that for the spinors $\la, \bar\la, \xi$ and $\bar\xi$ defined in \eqn{eqn:spinorchoice} above the off-shell information is chosen to reside in the $1\dot{1}$ (or $p_-$) component of the matrix \eqn{eqn:momentummatrix} completely. Thus, as long as one uses only $p_+,\,p_\bot$ and $\bar p_\bot$ for defining spinorial brackets, the resulting objects will have a precise meaning for off-shell vectors. In particular, we will use
\begin{equation}\label{eqn:spinorbrackets}
\langle p\, q\rangle\equiv\epsilon^{\alpha\beta}\lambda_{\alpha}\lambda_{\beta}=\sqrt{2}\frac{\left(p\, q\right)}{\sqrt{p_{+}}\,\sqrt{q_{+}}},\,\left[p\, q\right]\equiv\epsilon^{\dot{\alpha}\dot{\beta}}\bar{\lambda}_{\dot{\alpha}}\bar{\lambda}_{\dot{\beta}}=\sqrt{2}\left\{ p\, q\right\} \frac{\sqrt{p_{+}}\,\sqrt{q_{+}}}{p_{+}q_{+}}\,,
\end{equation}
where round and curly brackets are defined as
\begin{equation}
(p\, q)=p_+q_\bot-q_+p_\bot,\, \qquad \{ p\, q\} =p_+\bar{q}_\bot-q_+\bar{p}_\bot.
\end{equation}
In the above definition, the components $p_-$ and $q_-$ do not occur. For on-shell vectors the spinor brackets defined in \eqn{eqn:spinorbrackets} coincide with the usual definition. For a general non-lightlike vector they project onto its on-shell part (cf. \eqn{eqn:vectordecomposition}). Correspondingly, the usual identification of the scalar product and the spinor brackets via $2\,p\cdot q=\<p\, q\> [p\, q]$ is only valid as long as all participating vectors are on-shell, as can be shown by comparison with \eqn{eqn:scalarproduct}. Momentum conservation in terms of round and curly brackets reads
\begin{equation}\label{eqn:momentumconservation}
\sum_i\,(k\,i)=0\quad\text{and}\quad \sum_i\,\{k\,i\}=0\,.
\end{equation}


\subsection{Kinematical and dynamical supersymmetry on the lightcone}
\Nf SYM theory (\Ne supergravity) exhibits 16 (32) supersymmetries
\begin{equation}
\bar q_\ad^A\and q_{\a A}
\end{equation}
where $\a,\ad=1,2$ and $A=1,\ldots,4$ $(A=1,\ldots,8)$. The anticommutator of two supersymmetries yields (cf. \eqn{eqn:momentummatrix})
\begin{equation}\label{eqn:susycommutator}
\lbrace\bar q_\ad^A,q_{\a B}\rbrace=\delta^A_{\,\,B}\la_\a\lb_\ad.
\end{equation}
In the lightcone superfield actions for \Nf SYM and \Ne supergravity, \eqns{eqn:N4LCaction}{eqn:N8LCaction} below, which will be serving as starting point for our considerations, only half of the supersymmetries of either theory are manifest. Supersymmetries with generators $\bar q_{\dot 2}^A$ and $q_{2A}$ will be referred to as \textit{kinematical} supersymmetries as their anticommutators close onto the lightcone momentum $p_+$:
\begin{equation}\label{eqn:susykinem}
\lbrace\bar q_{\dot 2} ^A,q_{2 B}\rbrace=\delta^A_{\,\,B}p_{\dot 22}.
\end{equation}
Invariance of the actions \eqns{eqn:N4LCaction}{eqn:N8LCaction} under kinematical supersymmetry becomes obvious, once one identifies the generators $\bar q_{\dot 2}^A$ and $q_{2A}$ with the 8 (16) Grassmann coordinates $\theta^A$ and $\bar\theta_A$ of the lightcone superspace.

The remaining generators, $\bar q_{\dot 1}^A$ and $q_{1A}$ constitute the \textit{dynamical} supersymmetry. The name originates from the fact that their commutator is the lightcone Hamiltonian for the massless on-shell particles
\begin{equation}
\lbrace\bar q_{\dot 1}^A,q_{1 B}\rbrace=\delta^A_{\,\,B}\frac{p_\bot\bar p_\bot}{p_+}=\delta^A_{\,\,B}p_{\dot 1 1}.
\end{equation}
By splitting the supersymmetries in kinematical and dynamical ones and noticing that only the kinematical part is manifest, Lorentz covariance is broken.


\subsection{Lightcone vs. covariant  actions for \Nf SYM theory}
The starting point for the derivation of the novel path integral in \Nf SYM theory is the well-known lightcone Lagrangian of Brink, Nilsson and Lindgren \cite{Brink:1982pd}:
\begin{equation}\label{eqn:N4LCaction}
S^\Nf[\Phi, \bar \Phi] = S^2 + S^3 + S^4
\end{equation}
where
\begin{eqnarray}
S^2\!+\!S^3&=&\int d^{4}x\, d^{4}\theta\, d^{4}\bar{\theta}\, \left [ \bar{\Phi}^a \frac{\Box}{2 \, \pd^{+2}}\Phi^a  -\frac{2}{3} g f^{abc} \left(\frac{1}{\pd_{+}}\bar \Phi^a  \Phi^b\bar \pd {\Phi}^c +\frac{1}{\pd_{+}}{\Phi}^a \bar \Phi^b {\pd}\bar \Phi^c \right)\right ]
\nnl
S_{4}\!\!\!&=&\!\!\!-\frac{1}{2}g^{2}f^{abc}f^{ade}\int d^{4}x\, d^{4}\theta\, d^{4}\bar{\theta}\left[ \frac{1}{\pd_{+}}(\Phi^{b}\pd_{+}\Phi^{c})\frac{1}{\pd_{+}}(\bar{\Phi}^{d}\pd_{+}\bar{\Phi}^{e}) +{1\over 2}\Phi^{b}\bar{\Phi}^{c}\Phi^{d}\bar{\Phi}^{e}\right]\,.\nonumber
\label{S4}
\end{eqnarray}
Here small capital letters $a,\,b,\ldots$ denote color indices of the Yang-Mills gauge group $SU(N)$ with structure constants $f^{abc}$. The above lightcone Lagrangian depends on chiral and antichiral superfields $\Phi(x,\theta,\bar\theta)$ and $\bar\Phi(x,\theta,\bar\theta)$ and their lightcone derivatives. Upon integrating the action over $\theta\, , \bar{\theta}$, one recovers the component action for \Nf SYM theory in the $A^+=0$ gauge, as shown in \cite{Brink:1982pd}.
Due to the CPT-invariance of the \Nf supermultiplet the antichiral superfield $\bar\Phi(x,\theta^A,\bar\theta_A)$ is related to the chiral one via
\begin{equation}\label{eqn:N4duality}
\bar\Phi(x,\theta^A,\bar\theta_A)=-\frac{1}{4!}\pd_+^{-2}\eps^{ABCD}D_AD_BD_CD_D\Phi(x,\theta^A,\bar\theta_A),
\end{equation}
which suggests to remove one of them by virtue of the above equation. However, a particularly symmetric way of doing so is to introduce a new superfield $\phi$, which is related to the chiral as well as to the antichiral superfield by Fourier transformations, which in turn are chosen in concordance with \eqn{eqn:N4duality}. Two different forms of those transformation are available, both of which will be stated here, as either of them is used in the discussion of the preferred superframe in section \ref{sscn:amplitudes}.\par

Belitsky, Derkachov, Korchemsky and Manashov use the following transformation (just stating it for the field $\Phi$, cf. eqns (2.21) and (4.1) in \cite{Belitsky:2004sc})
\begin{equation}\label{eqn:BDKMtransform}
\Phi(x,\theta,\bar{\theta})=e^{\frac{1}{2}\bar{\theta}\cdot\theta\partial_{+}}\int\frac{d^{4}p}{(2\pi)^{4}}d^{4}\pi\, e^{ip\cdot x+\pi_A\theta^A}\phi_\text{BDKM}(p,\pi)
\end{equation}
where $p_i$ and $\pi_A, \,A=1,\ldots,4$ are the momentum associated with the particular superfield and the Grassmann coordinates of the chiral superspace respectively. Applying the transformation \eqn{eqn:BDKMtransform} to $S_3$ and $S_4$ leads to the BDKM form of the vertices 
\begin{equation}\label{eqn:BDKMvertices}
-2ig(2\pi)^4f^{abc}\left\{\{p_1\,p_2\}+\frac{(p_1\,p_2)\ki{p_1\,p_2}}{(p_{1+}p_{2+}p_{3+})^2}\right\}\delta^{(4)}\left(\sum_i \pi_{i}\right)\delta^{(4)}\left(\sum_i p_{i}\right)
\end{equation}
and the four-point contact term
\begin{eqnarray}\label{eqn:contactterms}
&&\Bigg\{ \! \left( f^{eab} f^{ecd} \frac{(p_1 -
p_2)_+ (p_3 - p_4)_+}{(p_1 + p_2)_+ (p_3 + p_4)_+} + f^{ead} f^{ebc} - f^{eac}
f^{edb} \right) \left( \frac{\ki{p_1, p_2}}{(p_{1+} p_{2+})^2} + \frac{\ki{p_3,
p_4}}{(p_{3+} p_{4+})^2} \right)
\nonumber\\
&& + \left( f^{eac} f^{edb} \frac{(p_1 - p_3)_+ (p_4 -
p_2)_+}{(p_1 + p_3)_+ (p_2 + p_4)_+} + f^{eab} f^{ecd} - f^{ead} f^{ebc} \right)
\left( \frac{\ki{p_2, p_4}}{(p_{2+} p_{4+})^2} + \frac{\ki{p_1, p_3}}{(p_{1+}
p_{3+})^2} \right)
\nonumber\\
&&+ \left( f^{ead} f^{ebc} \frac{(p_1 - p_4)_+ (p_2 -
p_3)_+}{(p_1 + p_4)_+ (p_2 + p_3)_+} + f^{eac} f^{edb} - f^{eab} f^{ecd} \right)
\left( \frac{\ki{p_1, p_4}}{(p_{1+} p_{4+})^2} + \frac{\ki{[p_2, p_3}}{(p_{2+}
p_{3+})^2} \right) \! \Bigg\}  \nnl
&&\quad\delta^{(4)}\left(\sum_i \pi_{i}\right)\delta^{(4)}\left(\sum_i p_{i}\right),
\end{eqnarray}
which will prove very useful below. Here 
\begin{equation}\label{eqn:fermbracket}
\ki{p_1\,p_2}=\prod\limits_{A=1}^{4}(\pi_{1,A}p_{2+}-\pi_{2,A}p_{1+})\quad\text{and satisfies}\quad\sum_i\,\ki{k\,i}=0.
\end{equation}
Fu and one of the authors employ the following definitions \cite{Fu:2010qi}:
\begin{equation}\label{eqn:Kalloshtransform1}
\Phi(x,\theta,\bar{\theta})=e^{\frac{1}{2}\bar{\theta}\cdot\theta\partial_{+}}\int\frac{d^{4}p}{(2\pi)^{4}}d^{4}\eta\, e^{ip\cdot x+\eta_A\frac{p_{+}}{\sqrt{p_{+}}}\theta^A}\,\left(\frac{-i}{p_{+}}\right)\phi(p,\eta)
\end{equation}
\begin{equation}\label{eqn:Kalloshtransform2}
\bar{\Phi}(x,\theta,\bar{\theta})=e^{-\frac{1}{2}\bar{\theta}\cdot\theta\partial_{+}}\int\frac{d^{4}p}{(2\pi)^{4}}d^{4}\eta\, e^{ip\cdot x}\,\delta^{4}(\bar{\theta}\sqrt{p_{+}}-i\eta)\left(\frac{-i}{p_{+}}\right)\phi(p,\eta).
\end{equation}
The Lie-algebra valued off-shell superfield $\phi(p, \eta)= \phi^a(p, \eta)  t^a$ is implicitely understood to be multiplied by the appropriate color-matrix $t^a$ of the Yang-Mills gauge group $SU(N)$. It depends on the momentum of the superparticle and can be expanded into
the physical degrees of freedom of the \Nf SYM theory with respect to the Grassmann variables $\e$:
\begin{equation}
 \phi  = A(p) + \e_A \psi ^A (p) + {1\over 2!} \e_{A} \e_B  \phi ^{AB}(p) + {1\over 3!} \eps ^{ABCD} \e_{A}\e_B \e_C  \bar \psi_D (p)+ {1\over 4!} \eps ^{ABCD} \e_{A}\e_B \e_C  \e_D  \bar A(p).
\label{PhiYM}
\end{equation}
Here $A(p)$ and $\bar A(p)$ are the positive and negative helicity gluon, $\psi^A$ and $\bar\psi_D$ the positive and negative helicity gluino and $\phi^{AB}$ a scalar field. Supersymmetry generators $\bar q$ and $q$ can be represented\footnote{Those definitions are in concordance with the ones chosen in \cite{Drummond:2008cr}.} in terms of $\la$ and $\e$ as
\begin{equation}\label{eqn:N4qs}
q_{A\a}=\la_\a\e_A,\and\bar q_\ad^A=\lb_\ad\frac{\pd}{\pd\e_A},
\end{equation}
where the versions applicable to a products of $n$ chiral superfields are given by
\begin{equation}\label{eqn:N4Qs}
 Q_{A\alpha}\equiv \sum_{i=1}^{n} \lambda_{i \alpha}  \eta_{i A} \, ,\qquad
\overline Q_{\dot \alpha}^A \equiv \sum_{i=1}^n \bar \lambda_{i \dot \alpha}{\partial \over \partial \eta_{Ai}}\,.
\end{equation}

Employing the transformations \eqns{eqn:Kalloshtransform1}{eqn:Kalloshtransform2}, cubic and quartic vertices have been derived from the Brink-Lindgren-Nilsson \cite{Brink:1982pd} form of the \Nf SYM lightcone action in \cite{Fu:2010qi}. The resulting lightcone action in chiral superspace can be written as
\begin{eqnarray}\label{eqn:N4action}
\CS^\Nf&=&\frac{1}{2}\int d^{8}z\varphi(z)p^2\varphi(-z)\nnl
&+&Cf^{abc}\int\prod\limits_{i=1}^{3}\{d^{8}z_i\varphi(z_i)\}
\left[\frac{\delta^4(\sum_ip_i)\delta^{8}(\sum_i\lambda^i\e_i)}{\<12\>\<23\>\<31\>}+
\frac{\delta^4(\sum_ip_i)\delta^{4}(\sfrac{1}{2}\ve^{ijk}[ij]\e_k)}{[12][23][31]}\right]\nnl
&+&\,\CS_4\,.
\end{eqnarray}
Here $C=\frac{g\cdot c_1c_2c_3(2\pi)^4}{3}$,
 where $c_i=sgn(p_{i+})$ and $d^8z_i=d^4pd^4\e/(2\pi)^4$.  In the above formula the second term corresponds to a $\overline{\text{MHV}}$-vertex, while the first is the MHV-part.

The contact term $\CS_4$ is given in equations (2.20) and (2.21) of reference \cite{Fu:2010qi}. It depends on $\psi_{ij}$, a related form of the fermionic brackets $\ki{\,}$ (see \eqn{eqn:defpsi} below for an definition of $\psi_{ij}$). Since we will be using the equivalent form of the counterterm derived in BDKM \eqn{eqn:contactterms}, we refrain from stating the lengthy expression here. In Appendix A we show the algebraic equivalence of the vertices in the new form of the action \eqn{eqn:N4action} with their BDKM counterparts \eqn{eqn:BDKMvertices}.

In order to be able to lateron connect the two approaches introduced above, the relation between the expressions and variables will be spelled out in the following lines. The superspace variable $\pi$ is related to $\eta$ via
\begin{equation}\label{eqn:fermrelation}
\pi_i=\eta\sqrt{p_{i+}}
\end{equation}
and the fields $\phi_\text{BDKM}$ and $\phi$ satisfy $\phi_\text{BDKM}=\left(\frac{-i}{p_+}\right)\phi$. The fermionic brackets \eqn{eqn:fermbracket} turn out to be closely connected to the antisymmetric objects
\begin{equation}\label{eqn:defpsi}
\psi_{ij}=\prod\limits_{A=1}^{4}(\eta_{i,A}\lb_{j\dot 2}-\eta_{j,A}\lb_{i\dot 2}),
\end{equation}
which have been defined in \cite{Fu:2010qi} and will show up as additional terms in the calculation of amplitudes in section \ref{scn:amplitudes} below. Explicitely, 
\begin{eqnarray}\label{eqn:fermbracketsresolve}
\ki{ij}&=&\prod\limits_{A=1}^{4}(\pi_{i,A}p_{j+}-\pi_{j,A}p_{i+})\nnl
&=&\prod\limits_{A=1}^{4}(\eta_{i,A}\sqrt{p_{i+}}p_{j+}-\eta_{j,A}\sqrt{p_{j+}}p_{i+})\nnl
&=&(p_{i+}p_{j+})^2\prod\limits_{A=1}^{4}(\eta_{i,A}\sqrt{p_{j+}}-\eta_{j,A}\sqrt{p_{i+}})\nnl
&=&(p_{i+}p_{j+})^2\,(\psi_{ij})_{\dot 2}\,,
\end{eqnarray}
where $\sqrt{p_+}$ has to be identified with the $\la_{\dot 2}$ component in order to preserve invariance under supersymmetry transformations. (see section \ref{scn:LCcov} for an explanation).


\subsection{Lightcone vs. covariant actions for \Ne supergravity}
The lightcone action for \Ne supergravity in real superspace derived\footnote{An earlier though more complicated version of the action was found in \cite{Bengtsson:1983pg}.}
 in \cite{Ananth:2005vg,Brink:2008qc} reads
\begin{equation}\label{eqn:N8LCaction}
S^\Ne[\Phi, \bar \Phi] = \int d^4x\, d^8\theta\, d^8\bar\theta\, \left [ \bar\Phi\frac{\Box}{\pd^4_+}\Phi  +\frac{3\k}{2}\left(\frac{1}{\pd_+^2}\bar\Phi\bar\pd\Phi\bar\pd\Phi +\frac{1}{\pd_+^2}\Phi\pd\bar\Phi\pd\bar\Phi\right)+\CO(\k^2)\right ]\,,
\end{equation}
where the chiral superfield $\Phi(x,\theta,\bar\theta)$ and its antichiral counterpart $\bar\Phi(x,\theta,\bar\theta)$ $(\theta,\bar\theta=1\ldots 8)$) are understood to include one power of the gravitational coupling constant $\k$. They are again related due to the CPT invariance of the \Ne supergravity supermultiplet:
\begin{equation}
\bar\Phi(x,\theta,\bar\theta)=-\frac{1}{8!}\pd_+^{-4}\eps^{ABCDEFGH}D_AD_BD_CD_DD_ED_FD_GD_H\Phi(x,\theta,\bar\theta).
\end{equation}
After integration over $\theta$ and $\bar\theta$ one should recover the supersymmetric extension of the Einstein-Hilbert Lagrangian in lightcone gauge $g_{--}=g_{-i}=0$.
In parallel to the situation in \Nf SYM theory the action can be expressed in terms of one superfield $\varphi(p,\e)$, $\e_A: A=1\ldots 8$ via
\begin{eqnarray}
\Phi(x,\theta,\bar{\theta})&=&e^{\frac{1}{2}\bar{\theta}\cdot\theta\partial_{+}}\int\frac{d^4p}{(2\pi)^4}d^8\e\, e^{ip\cdot x+\eta_A\frac{p_{+}}{\sqrt{p_{+}}}\theta^A}\,\left(\frac{-i}{(p_+)^2}\right)\varphi(p,\eta)\nnl
\bar{\Phi}(x,\theta,\bar{\theta})&=&e^{-\frac{1}{2}\bar{\theta}\cdot\theta\partial_{+}}\int\frac{d^{4}p}{(2\pi)^4}d^8\e\, e^{ip\cdot x}\,\delta^8(\bar{\theta}\sqrt{p_{+}}-i\eta)\left(\frac{-i}{(p_+)^2}\right)\varphi(p,\eta).
\end{eqnarray}
Depending on eight Grassmannian variables $\e$, the superfield $\varphi(p,\e)$ can be expanded into physical degrees of freedom
\begin{eqnarray}\label{eqn:expansion}
\varphi(p,\e)&=&\bar h(p)+\e_A\psi^A(p)+\e_{AB}B^{AB}(p)+\e_{ABC}\chi^{ABC}(p)+\e_{ABCD}\phi^{ABCD}(p)\nnl
&&+\tilde\e^{ABC}\tilde\chi_{ABC}(p)+\tilde\e^{AB}\tilde B_{AB}(p)+\tilde\e^A\tilde\psi_A(p)+\tilde\e h(p)\,,
\end{eqnarray}
where $\e_{A_1\ldots A_n}=\sfrac{1}{n!}\e_{A_1}\ldots\e_{A_n}$ and $\e^{A_1\ldots A_n}=\eps^{A_1\ldots A_nB_1\ldots B_{n-8}}\e_{B_1}\ldots\e_{B_{n-8}}$\,.
In complete analogy to the \Nf case and the coventions in \cite{Drummond:2009ge} one can represent supersymmetry generators as
\begin{equation}\label{eqn:N8qs}
q_{A\a}=\la_\a\e_A,\quad\bar q_\ad^A=\lb_\ad\frac{\pd}{\pd\e_A},\quad Q_{A\alpha}\equiv \sum_{i=1}^{n} \lambda_{i \alpha}  \eta_{i A} \, ,\quad\text{and}\quad
\overline Q_{\dot \alpha}^A \equiv \sum_{i=1}^n \bar \lambda_{i \dot \alpha}{\partial \over \partial \eta_{Ai}}\,.
\end{equation}

The terms from the lightcone action \eqn{eqn:N8LCaction} translate again into a kinetic term and two three-point interactions (MHV and $\overline{\text{MHV}}$). In addition, there will be possibly an infinite number of contact terms which originate from terms of $\CO(\k^2)$. As an example, the calculation for the three-point MHV-vertex will be sketched here, where the $\overline{\text{MHV}}$ vertex and the kinematic term can be derived using the same methods.

Starting with
\begin{eqnarray}
\CS_3&=&\frac{3\k}{2}\int d^4xd^8\theta d^8\bar\theta\,\frac{1}{\pd_+^2}\bar\Phi\bar\pd\Phi\bar\pd\Phi\nnl
   &=&\frac{3\k}{2}\int d^4xd^8\theta d^8\bar\theta\prod\limits_{i=1}^{3}\{d^{12}z_i\varphi(z_i)\}\left(\frac{-i}{(p_{1+})^2}\right)\!\left(\frac{-i}{(p_{2+})^2}\right)\!\left(\frac{-i}{(p_{3+})^2}\right)\!\left(\frac{(p_{2\bot}-p_{3\bot})^2}{(p_{1+})^2}\right)\!\times\nnl
&&\quad e^{i\sum_ip_i\cdot x}e^{\sfrac{1}{2}\bar\theta\cdot\theta(p_{1+}-p_{2+}-p_{3+})}e^{\e_1\frac{p_{1+}}{\sqrt{p_{1+}}}\theta}\delta^8(\bar\theta\sqrt{p_{2+}}-i\e_2)\delta^8(\bar\theta\sqrt{p_{3+}}-i\e_3)\,,
\end{eqnarray}
where $d^{12}z_i=d^4pd^8\e/(2\pi)^4$, integration over $d^4x$ will produce the momentum conserving $\delta$-function, whose application renders the last line into
\begin{equation}
\delta^4(\sum_ip_i)e^{i\theta(\bar\theta p_{1+}-i\e_1\frac{p_{1+}}{\sqrt{p_{1+}}})}\delta^8(\bar\theta\sqrt{p_{2+}}-i\e_2)\delta^8(\bar\theta\sqrt{p_{3+}}-i\e_3)\,.
\end{equation}
Performing the integration over $d^8\theta$ as a next step and pulling out some factors of $p_+$, the last line reads now
\begin{equation}
\delta^4(\sum_ip_i)(p_{1+})^8(p_{2+}p_{3+})^4\delta^8(\bar{\theta}-i\eta_{1}/\sqrt{p_{1+}})\delta^8(\bar{\theta}-i\eta_2/\sqrt{p_{2+}})\delta^8(\bar{\theta}-i\eta_3/\sqrt{p_{3+}}).
\end{equation}
Focussing on the $\bar\theta$-integration
\begin{equation}
\int d^{4}\bar\theta\,\delta^8(\bar{\theta}-i\eta_{1}/\sqrt{p_{1+}})\delta^8(\bar{\theta}-i\eta_2/\sqrt{p_{2+}})\delta^8(\bar{\theta}-i\eta_3/\sqrt{p_{3+}}),
\end{equation}
one can show that the three $\delta$-functions finally yield the expected supermomentum conservation
\begin{equation}
=\prod_{A=1}^8\frac{\eta_{1A}\eta_{2A}}{\sqrt{p_{1+}}\sqrt{p_{2+}}}+\frac{\eta_{2A}\eta_{3A}}{\sqrt{p_{2+}}\sqrt{p_{3+}}}+\frac{\eta_{3A}\eta_{1A}}{\sqrt{p_{3+}}\sqrt{p_{1+}}}
=\frac{1}{(\sqrt{2})^8(12)^{8}}\prod_{A=1}^{8}\sum_{ij}\left\langle ij\right\rangle \eta_{iA}\eta_{jA}
\end{equation}
by employing relations \eqn{eqn:spinorbrackets} and noting that for a three-point interaction $(12)=(23)=(31)$ (cf. \eqn{eqn:momentumconservation}). The product of sums in the above equation, however, is nothing as the supermomentum conserving $\delta$-function $\delta^{16}(\sum_i\lambda^i\e_i)$. Thus the resulting expression reads
\begin{eqnarray}
\!\!\!\!\!\!\CS_3\!\!&\!\!=\!\!&\!\!\frac{3 i \k}{2}\prod\limits_{i=1}^{3}\{d^{12}z_i\varphi(z_i)\}\delta^4(\sum_ip_i)\delta^{16}(\sum_i\lambda^i\e_i)(p_{1+}p_{2+}p_{3+})^2\frac{(p_{1+}p_{2\bot}\!-\!p_{1+}p_{3\bot})^2}{(\sqrt{2})^8(12)^8}\,.
\end{eqnarray}
In the path integral the vertex will be contracted with three totally symmetric scalar superfields. Therefore the last term should be symmetrized. Since $(p_{1+}p_{2\bot}-p_{1+}p_{3\bot})=(32)-p_{3+}p_{3\bot}+p_{2+}p_{2\bot}$ and symmetrization will remove all terms of the form $p_{i+}p_{i\bot}$, one is left with
\begin{equation}
\frac{(p_{1+}p_{2\bot}-p_{1+}p_{3\bot})^2}{(\sqrt{2})^8(12)^8}=
\frac{(23)^2}{(\sqrt{2})^8(12)^8}=\frac{1}{\sqrt{2}^8(12)^6}\,.
\end{equation}
After restoring the original spinor-helicity brackets in the denominator employing \eqn{eqn:spinorbrackets}, one finally finds
\begin{eqnarray}
\CS_3&=&\frac{3 i \k}{4}\prod\limits_{i=1}^{3}\{d^{12}z_i\varphi(z_i)\}\frac{\delta^4\left(\sum_ip_i\right)\delta^{16}(\sum_i\lambda^i\e_i)}{(\<12\>\<23\>\<31\>)^2}\,.
\end{eqnarray}
Repeating the same procedure for the kinetic term and the $\overline{\text{MHV}}$-vertex, one finds for the symmetric lightcone action for \Ne supergravity
\begin{eqnarray}\label{eqn:N8action} 
\CS^\Ne&=&\frac{1}{2}\int d^{12}z\varphi(z)p^2\varphi(-z)\nnl
&+&\frac{3 i\k}{4}\int\prod\limits_{i=1}^{3}\{d^{12}z_i\varphi(z_i)\}
\left[\frac{\delta^4(\sum_ip_i)\delta^{16}(\sum_i\lambda^i\e_i)}{(\<12\>\<23\>\<31\>)^2}+
\frac{\delta^4(\sum_ip_i)\delta^{8}(\sfrac{1}{2}\ve^{ijk}[ij]\e_ks)}{([12][23][31])^2}\right]\nnl
&+&\CO(\k^2) \,.
\end{eqnarray}


\subsubsection{Off-shell unbroken supersymmetry of the quadratic part of the action}
Starting from \Ne supergravity action \eqn{eqn:N8action}, one can rewrite the quadratic part as
\be
\CS^\Ne_2 = {1\over 2}  \int \prod _{i=1}^{2} \{d^{12} z_i\,  \varphi (z_i)\} \delta^4( p_1+p_2) \delta^8( \eta_1+\eta_2) \, p_1^2
\label{S2}\,,\ee
where supermomentum conservation appears after employing $\la(p_1) = \la(p_2)=\la(-p_1)$ (see \eqn{eqn:lachange} above). In this form it is easy to show that the quadratic part of the action has 32 unbroken supersymmetries despite the superfield $\varphi(z_i)$ is off-shell. 

Explicitely, $Q_{A\alpha}=\la_\a^1\eta_{1A} + \la_\a^2\eta_{2A}=\la^1_\a(\e_1+\e_2)$ and thus $Q_{A\alpha}\CS_2=0$ by virtue of supermomentum conservation. Acting with the other supersymmetry generator $\overline Q^A_{\dot\alpha}=\left (\bar \la^1 {\pd \over \pd \eta_{A1}}  + \bar \la^2 {\pd \over \pd \eta_{A2}} \right)$ on $\d^8( \eta_1+\eta_2)$ leads to $\bar\la_\a^1+\bar\la_\a^2$, which vanishes by use of \eqn{eqn:lachange}.


\subsubsection{Off-shell broken dynamical supersymmetry of cubic vertices}\label{ssn:susbreak}
While non of the supersymmetries is broken for the kinetic term, this is not true for the three-vertices. While acting with $Q_{A \alpha}= \sum_{i=1}^3\la_\a^i\eta_{1A}$ on the cubic MHV part 
\begin{equation}
\CS_3^\Ne=\int\prod\limits_{i=1}^{3}\{d^{12}z_i\varphi(z_i)\}
\frac{\delta^4(\sum_ip_i)\delta^{16}(\sum_i\lambda^i\e_i)}{(\<12\>\<23\>\<31\>)^2} \label{1S3}
\end{equation}
yields zero, this is not true for the barred set of SUSY generators because
\begin{equation}
 \overline Q_{\dot \alpha}^A   (\sum_{i=1}^3 \lambda_{\alpha i} \eta_{i B})= 
\sum_{j=1}^3 \bar \la_{j \dot \a}{\pd \over \pd \e_{Ai}} \left(\sum_{i=1}^3 \la_{\a i} \e_{i B}\right)=
\sum_{i=1}^3   \bar{\la}_{\dot \a} \la_\a \, \d^A{}_B\,.
\end{equation}
Unfortunately momentum conservation does not read $\sum_{i=1}^3 \bar \lambda_{\dot \alpha i}  \lambda_{\alpha i} =0$ because the off-shell contributions have to be taken into account. In particular, $\delta^{(4)}(\sum_i p_{i})$ implies
\begin{equation}\label{eqn:offshellpart}
\sum_{i=1}^3 \bar \lambda_{\dot \alpha} \lambda_{\alpha i} = - \sum_{i=1}^3 \bar \xi_{\dot \alpha} \xi_{\alpha i} \, ,
\qquad
 \sum_{i=1}^3 \bar \xi_{\dot 1} \xi_{1 i}=  \sum_{i=1}^3 {p_i^2\over \sqrt{2} \, p_{i+}}\,.
\end{equation}
Thus this part of the action is not invariant under the dynamical supersymmetries $ \overline Q_{\dot 1}^A$.

A similar situation is encountered for the three-point $\overline{\text{MHV}}$ vertex of the action, where now the roles of the supersymmetry operators are exchanged. Acting with $\epsilon^{ \dot \alpha}_A  \overline Q^A_{\dot \alpha} $ onto
\begin{equation}\label{eqn:MHVBAR}
 \int \prod_{i=1}^{i=3} \{d^{12} z_i  {\phi}(z_{i})\}
{ \d^{(4)}(\sum_i p_{i})\d^8 \left(\left[12\right]\eta_{3}+\left[23\right]\eta_{1}+\left[31\right]\eta_{2}\right)\over (\left[12\right]\left[23\right]\left[31\right])^2}
\end{equation}
produces an expression which vanishes due to Schouten identity without employing momentum conservation:
\begin{equation}
[\epsilon 1][23] + [\epsilon 2][31] + [\epsilon 3][12]=0\,.
\end{equation}
On the other hand, the action of $Q_{A\alpha}$ onto \eqn{eqn:MHVBAR} yields 
\begin{equation}
 \sum_{i=1}^3 \lambda_\alpha (p_i) \eta_{i } = {\lambda_\alpha (p_1) \over [23]} \left(\left[12\right]\eta_{3}+\left[23\right]\eta_{1}+\left[31\right]\eta_{2}\right) +  \sum_{i=1}^3   \lambda_\alpha \bar \lambda _{\dot \alpha} (p_i) \, {\bar \lambda^{\dot  \alpha} (p_3) \eta_2 -\bar  \lambda^{\dot \alpha} (p_2) \eta_3 \over [23]}\,.
 \end{equation}
While the first term vanishes on the support of $\delta^8 \left(\left[12\right]\eta_{3}+\left[23\right]\eta_{1}+\left[31\right]\eta_{2}\right)$, the second term is proportional to  $ \sum_{i=1}^3  \xi_\alpha  \bar{ \xi} _{\dot \alpha} (p_i)= \sum_{i=1}^3 {p^2_i\over \sqrt 2 \, p_{i+}}$ (cf. \eqn{eqn:offshellpart}). Thus dynamical supersymmetry with operators $Q_{A 1}$ is broken.

In both cases above the terms deviating from dynamical supersymmetry take the same form at each vertex: 
\be
\delta_{\bar \epsilon^{\dot 1}_A} {\cal S}^3_{ {MHV}}\sim \sum_{i=1}^3 {p^2_i\over p_{i+}}   \, , \qquad  \qquad \delta_{ \epsilon^{ 1 A} } {\cal S}^3_{\overline {MHV}} \sim \sum_{i=1}^3 {p^2_i\over p_{i+}}\,.
\ee
Which role do those supersymmetry breaking terms play in the evaluation of Feynman diagrams? As an example, consider a fourpoint diagram with two three-point vertices and one internal propagator
\begin{equation}
\psset{xunit=6pt,yunit=6pt,linewidth=0.7pt}
\begin{pspicture}(8,5)
\psline{-}(0,0)(2.5,2.5)\psline{-}(0,5)(2.5,2.5)\psline{-}(2.5,2.5)(5.5,2.5)\psline{-}(5.5,2.5)(8,0)\psline{-}(5.5,2.5)(8,5)
\rput[b](5.2,0.7){\small{j}}\rput[b](2.7,0.8){\small{i}}
\end{pspicture}\,.
\end{equation}

All external legs being assumed to be on-shell the only fields whose off-shell part we have to consider are those attached to the internal propagator. Explicitely, the most divergent terms are of the form
\begin{equation}
{p_i^2}  T\left (\phi(p_i, \eta_i) \phi(p_j, \eta_j) \right) \delta^4(p_i+ p_j) \sim {p_i^2} \,{1\over p_i^2} \sim 1\,,
\label{shrink}
\end{equation}
which correspond to 
\begin{equation}
{\Box_x } T\left (\phi(x ) \phi(y) \right) \sim   \delta^4(x-y)
\label{shrink1} 
\end{equation}
in coordinate space. The Lorentz covariant propagator shrinks to a point and the corresponding term exhibits the structure of a contact term. In other words, the terms breaking dynamical supersymmetry are at the same level as the (unknown) contact term interactions. If one however deforms the momenta in a way as to set the internal propagator on-shell, full supersymmetry is restored. 


\section{From the lightcone to covariant amplitudes}\label{scn:LCcov}

On the way from the lightcone path integrals \eqns{eqn:N4LCaction}{eqn:N8LCaction} to the covariant expressions given in \eqns{eqn:N4action}{eqn:N8action} and during the amplitude calculations below, several issues related to complexification, choice of signature and the identification of $p_+$ with different components of $\la$ and $\lb$ arise. This section discusses those problems and explains our approach.

\paragraph{Three-point vertices} As explained in numerous places, any three vertex configuration of three real on-shell momenta in Minkowski signature vanishes trivially. As long as one stays at the tree level, there are two possible resolutions: in a setup with complexified momenta one can either choose all $\la$ or all $\lb$ to be zero, which allows either a nonzero $\overline{\text{MHV}}$ or MHV amplitude respectively. Another possibility is to perform the calculation in Kleinian signature $(--++)$, where one can find a nontrivial kinematical configuration. Once loop amplitudes shall be calculated one will be concerned with unitary cuts, which do not have a clear meaning in Kleinian signature \cite{Britto:2004nc}. However, here we will be employed with tree amplitudes only and take an operational approach by interpreting the spinor brackets - once they have been obtained in a complexified geometry - in Minkowski space.

\paragraph{Chiral superspace} Although showing up as a shortcoming of the formulation used, complex momenta are a natural feature of chiral superspace. In the simplest case of $\CN=1$ supersymmetry the chiral basis is obtained by performing a complex shift from the real coordinate $x^\mu$ to the complex $y^\mu= x^\mu +i \theta \sigma^\mu \bar \theta$. Correspondingly the chiral superfield $\Phi(x, \theta, \bar \theta) $, $\bar D_{\dot \a} \Phi=0$ does not depend on $\bar\theta$ after the shift: $\Phi=\Phi(y, \theta)$. Since the coordinate $y^\mu$ is complex, its Fourier transformation $p^\mu$ is naturally complex as well. Those considerations work in exactly the same way for the lightcone scalar superfield in the chiral basis.

\paragraph{Which of the $\la,\lb$ shall be identified with $\sqrt{p_+}\,$?} Considering \eqn{eqn:spinorchoice}, $\sqrt{p_+}$ is identified with two distinct spinor-comonents, $\la_2$ and $\lb_{\dot 2}$, simultaneously. However, once the complexification or Kleinian signature has been agreed on, $\la_2$ and $\lb_{\dot 2}$ can be treated as independent variables and are not linked by \eqn{eqn:spinorchoice} any more. In other words: in a complexified setup a transformation is allowed to change $\la_2$ without affecting $\lb_{\dot 2}$ and vice versa. In particular, the vanishing of one of them does not require the vanishing of the other one.
With this prescription one additional comment is in place: in order to make the lightcone momentum component
\begin{equation}
p_+= p_{2\dot 2}= \la_2(p) \bar\la_{\dot 2}(p)
\end{equation}
vanish, it is now sufficient to have just \textit{one} of the independent spinors vanishing. The analysis in the sections below uses this fact extensively.

Nevertheless, even with the above statement it is not clear, which of $\la_2$ and $\lb_{\dot 2}$ shall be identified with $\sqrt{p_+}$ in different expressions. Fortunately, supersymmetry provides a sufficient argument: each term in the lightcone action must be invariant under the kinematical supersymmetry operators $Q_2$ and $\bar Q_{\dot 2}$ defined in \eqns{eqn:N4Qs}{eqn:N8qs}.
The first questionable term is
\begin{equation}
  \sum _i \pi_i= \sum_i \sqrt {p_+^i } \eta_i.
\end{equation}
In order for the complete expression to be invariant under supersymmetry, one has to identify $\sqrt {p_+^i }$ with $\la_2^i$. The resulting expression yields then $\sum_i \la_2^i \eta_i= Q_2$.
On the contrary, in the fermionic brackets \eqn{eqn:fermbracketsresolve}
\begin{eqnarray}
\ki{ij}=(p_{i+}p_{j+})^2\prod\limits_{A=1}^{4}(\eta_{i,A}\sqrt{p_{j+}}-\eta_{j,A}\sqrt{p_{i+}})
= (p_{i+}p_{j+})^2\prod\limits_{A=1}^{4} (\psi_{ij})_{\dot 2}
\end{eqnarray}
the $\sqrt{p_+}$ in the product has to be identified with $\bar \la_{\dot 2}$, which makes the expression invariant under $\bar Q_{\dot 2}$: $\bar Q_{\dot 2}(\psi_{ij})_{\dot 2}=0$. The other possible choice, $(\psi_{ij})_2$ would lead to $\bar Q_{\dot 2}(\psi_{ij})_{2}\neq 0$.
With the above choice, one can nicely connect the lightcone formulation of BDKM to the Lorentz covariant formulation in \cite{Drummond:2008vq}: in the momentum superspace the lightcone supersymmetry generator $\pi= \sqrt p_+ \eta$ in \eqn{eqn:fermrelation} is promoted to $q_2=\la_2 (p) \eta$. Correspondingly, $\bar q_{\dot 2}$ becomes $\bar \la_{\dot 2} (p) {\partial \over \partial \eta}$  \cite{Kallosh:2009db,Fu:2010qi} (cf. \eqns{eqn:N4qs}{eqn:N8qs}).


\section{Amplitudes in \Nf SYM}\label{scn:amplitudes}


\subsection{\Nf SYM 4-point amplitude in a preferred superframe}\label{sscn:amplitudes}
Using the \Nf lightcone action found above, a first test of the formalism is the calculation of the fourpoint function. The following diagrams will contribute, where a circled vertex corresponds to $\overline{\text{MHV}}$, and an uncircled one to $\text{MHV}$.
\begin{equation}\label{eqn:fourpointpicture}
 \psset{xunit=4pt,yunit=4pt,linewidth=0.7pt}
 \begin{pspicture}(90,14)
 \rput[l](0,8)
 {
\psline{-}(0,0)(2.5,2.5)\psline{-}(0,5)(2.5,2.5)\psline{-}(2.5,2.5)(5.5,2.5)\psline{-}(5.5,2.5)(8,0)\psline{-}(5.5,2.5)(8,5)
\rput[b](-1,4.2){\small{1}}\rput[b](9,4.2){\small{2}}\rput[b](9,-1){\small{3}}\rput[b](-1,-1){\small{4}}
\pscircle(5.5,2.5){0.08}
\rput[b](4,-7.5){\small{\textbf{1}}}\pscircle(4,-6.6){0.3}
}
 \rput[l](14,8)
 {
\psline{-}(0,0)(2.5,2.5)\psline{-}(0,5)(2.5,2.5)\psline{-}(2.5,2.5)(5.5,2.5)\psline{-}(5.5,2.5)(8,0)\psline{-}(5.5,2.5)(8,5)
\rput[b](-1,4.2){\small{1}}\rput[b](9,4.2){\small{2}}\rput[b](9,-1){\small{3}}\rput[b](-1,-1){\small{4}}
\pscircle(2.5,2.5){0.08}
\rput[b](4,-7.5){\small{\textbf{2}}}\pscircle(4,-6.6){0.3}
}
 \rput[l](28,6.5)
 {
\psline{-}(0,0)(2.5,2.5)\psline{-}(5,0)(2.5,2.5)\psline{-}(0,8)(2.5,5.5)\psline{-}(5,8)(2.5,5.5)\psline{-}(2.5,5.5)(2.5,2.5)
\rput[b](-1,7.2){\small{1}}\rput[b](6,7.2){\small{2}}\rput[b](6,-1){\small{3}}\rput[b](-1,-1){\small{4}}
\pscircle(2.5,2.5){0.08}
\rput[b](2.5,-6){\small{\textbf{3}}}\pscircle(2.5,-5.1){0.3}
}
\rput[l](42,6.5)
 {
\psline{-}(0,0)(2.5,2.5)\psline{-}(5,0)(2.5,2.5)\psline{-}(0,8)(2.5,5.5)\psline{-}(5,8)(2.5,5.5)\psline{-}(2.5,5.5)(2.5,2.5)
\rput[b](-1,7.2){\small{1}}\rput[b](6,7.2){\small{2}}\rput[b](6,-1){\small{3}}\rput[b](-1,-1){\small{4}}
\pscircle(2.5,5.5){0.08}
\rput[b](2.5,-6){\small{\textbf{4}}}\pscircle(2.5,-5.1){0.3}
}
 \rput[l](56,8)
 {
\psline{-}(0,0)(5.5,2.5)\psline{-}(0,5)(2.5,2.5)\psline{-}(2.5,2.5)(5.5,2.5)\psline{-}(5.5,2.5)(8,5)
\psline{-}(8,0)(4.5,1.54)
\psline{-}(2.5,2.5)(3.5,2.09)
\rput[b](-1,4.2){\small{1}}\rput[b](9,4.2){\small{2}}\rput[b](9,-1){\small{3}}\rput[b](-1,-1){\small{4}}
\pscircle(5.5,2.5){0.08}
\rput[b](4,-7.5){\small{\textbf{5}}}\pscircle(4,-6.6){0.3}
}
 \rput[l](70,8)
 {
\psline{-}(0,0)(5.5,2.5)\psline{-}(0,5)(2.5,2.5)\psline{-}(2.5,2.5)(5.5,2.5)\psline{-}(5.5,2.5)(8,5)
\psline{-}(8,0)(4.5,1.54)
\psline{-}(2.5,2.5)(3.5,2.09)
\rput[b](-1,4.2){\small{1}}\rput[b](9,4.2){\small{2}}\rput[b](9,-1){\small{3}}\rput[b](-1,-1){\small{4}}
\pscircle(2.5,2.5){0.08}
\rput[b](4,-7.5){\small{\textbf{6}}}\pscircle(4,-6.6){0.3}
}
 \rput[l](84,8)
 {
\psline{-}(0,0)(5,5)\psline{-}(0,5)(5,0)
\rput[b](-1,4.2){\small{1}}\rput[b](6,4.2){\small{2}}\rput[b](6,-1){\small{3}}\rput[b](-1,-1){\small{4}}
\rput[b](2.5,-7.5){\small{\textbf{7}}}\pscircle(2.5,-6.6){0.3}
}
 \end{pspicture}
\end{equation}
In order to see the formalism at work, a calculation of diagram $1$ has been performed in \cite{Fu:2010qi} where further details can be found. Starting from the expression
\small
\begin{equation}
\int d^4P d^4\eta_P\,{\delta^4(p_1+p_4+P) \delta^8 (\la^1\eta_1+ \la^4 \eta_4 + \la^P \eta_P ) \delta^4(-P+p_2+p_3)  \delta^4( \eta_P [23] + \eta_2[3P] +\eta_3 [P2])\over P^2 \<41\>[23] \< 1P\> \< P4\> [P2] [3P] }
\label{BHT1}
\end{equation}
\normalsize
the integrations over $P$ and $\eta_P$ yielded
\begin{equation}\label{eqn:fourpointresult}
\frac{ \delta^4\left(\sum_{i=1}p_i\right)\delta^4 \left (\sum _{i=1}^4 \la^i_2 \eta_i  \right) \delta^4\left (\sum _{i=1}^4 \la^i_1 \eta_i -{\eta_2 ( \xi^P \bar \xi^P \bar \la^3)_1  - \eta_3 ( \xi^P \bar \xi^P \bar \la^2)_1 \over [23]} \right)}
{\<23\>\<41\>(\<12\>-\Delta_{13})
( \<34\>-\Delta_{42})}.
\end{equation}
The known Parke-Taylor result
\begin{equation}
\frac{\delta^4\left(\sum_{i=1}p_i\right)\delta^4 \left (\sum _{i=1}^4 \la^i_2 \eta_i  \right) \delta^4\left (\sum _{i=1}^4 \la^i_1 \eta_i \right)}{\<23\>\<41\>\<12\>\<34\>}
\end{equation}
is accompanied by additional terms proportional to the off-shell portion of the internal momentum $P$.
Having a look to the numerator first, the additional term in the $\la_1^i\eta_i$-part of the supermomentum conservation can be rewritten as
\begin{equation}
-{\eta_2 ( \xi^P \bar \xi^P \bar \la^3)_1  - \eta_3 ( \xi^P \bar \xi^P \bar \la^2)_1 \over [23]}
=-(\xi^P \bar \xi^P)_{1\dot{1}}\frac{\eta_2\bar\la^3_{\dot 2}-\eta_3\bar\la^2_{\dot 2}}{[23]}
=-(\xi^P \bar \xi^P)_{1\dot{1}}\frac{(\psi_{23})_{\dot 2}}{[23]}\,
\label{psi}\end{equation}
where $\psi_{ij}$ was defined in \eqn{eqn:defpsi}. In the denominator the terms $\Delta_{ij}$ are
\begin{equation}\label{Delta}
\Delta_{ij}=\frac{\la^i_2(\xi^P\bar\xi^P)_{1\dot 1}\bar\la^j_{\dot 2}}{[23]}\,.
\end{equation}
Those terms are expected to cancel for the on-shell amplitude after calculating all contributing diagrams including the contact terms: although Lorentz invariance is not manifest in the lightcone formalism, the on-shell amplitudes derived from the path integral corresponding to the action \eqn{eqn:N4action} should certainly be Lorentz invariant. Since proving these cancellations is algebraically involved, it is easier to pursue another direction, as will be elaborated on below.

Noticing the unwanted terms to be proportional to the off-shell part of the internal momentum, this result begs for a BCFW-shift in order to set the internal propagator on-shell, thus removing all unwanted terms at once. However, in order for BCFW to be valid, amplitudes should vanish for large values of the shifting parameter $z$. In addition, the final expression must not depend on the shifting parameter. This has been demonstrated for the maximally supersymmetric theories based on considering color-ordered subamplitudes (see \cite{ArkaniHamed:2008yf} for example). As these subamplitudes receive contributions from contact terms, which are known to not vanish at large $z$, cancellations between nonvanishing-$z$-parts of different diagrams have to take place. In the approach described below, each contributing diagram (up to the one delivering the solution) will vanish separately.  

One property to recall from the BCFW formalism is the following: a particular complex shift sets the internal propagator of \textit{one} of the contributing diagrams on-shell and also makes a selection about which of the vertices is MHV and which one is $\overline{\text{MHV}}$. This is done such that after the complex deformation in any other than the selected diagram the momentum- and supermomentum conserving $\delta$-functions can not be satisfied simultaneously.
Getting back to the result \eqn{eqn:fourpointresult}, the approach is different in the sense that here no specific complex deformation of the external momenta shall be performed, but it will be shown that in the lightcone formalism the choice of a particular superframe is sufficient to exclude all other diagrams including contact terms and thus leads to the correct result.

The procedure is reminiscent of the educated choice of reference momenta for a pure-gluon four-point amplitude in \cite{Dixon:1996wi}. Therein for each of the external gluon lines a reference momentum was chosen in a way to leave only one diagram and no contact term contributing to the color-ordered amplitude $A(1,2,3,4)$. With another choice of reference momenta for the external legs one can show the vanishing of any gluon-amplitude of the form $(++\cdots++)$ and $(-+\cdots++)$. This result can be derived easily by supersymmetric methods and is related to the secret supersymmetry of the pure-gluon QCD amplitudes. Since there is some freedom in choosing a superspace frame, it is an obvious idea to use a the freedom in the supervariables to simplify the calculation of the lightcone four-point amplitude. 

Let us repeat the logic: starting from the non-manifestly Lorentz-invariant lightcone formulation, the result of \textit{one} Feynman graph calculation is found to yield the expected result and additional terms which will add up with contributions from other graphs to a Lorentz invariant result in an algebraically involved way. We will show below that one can make the additional terms disappear by choosing a particular superframe. With this choice, no other diagrams including the contact term contribute to the calculation.


\subsubsection{Choice of a Lorentz frame}

Before getting to the superpart of the frame, consider the Lorentz-part of a superframe. We would like to find a large Lorentz transformation which can make the Lorentz non-covariant terms terms in (\ref{Delta})
\begin{equation}
\la^1_2(\xi\bar\xi)_{1\dot 1}\la^3_{\dot 2}\and\la^4_2(\xi\bar\xi)_{1\dot 1}\la^2_{\dot 2}
\end{equation}
disappear by setting components of certain spinors to zero. The off-shell component $\xi\bar\xi_{1\dot{1}}$ cannot be altered by a Lorentz transformation as it is proportional to the total momentum squared, which is a Lorentz scalar. 

In Minkovskian signature, the matrices $\Lambda_\a^\b$ and $\bar\Lambda_\ad^\bd$ representing a Lorentz tranformation on the right- and left-handed spinors should be complex conjugates. In order to remove the terms $\Delta_{13}$ and $\Delta_{42}$ one should in each of them set either $\la$ or a $\bar\la$ to zero. Considering a rotation $\Lambda\in SL(2,\ZC)$ for the unbarred $\la$'s, it is not possible to set two of them to zero simultaneously, so one has to employ a second rotation $\bar\Lambda\in SL(2,\ZC)$ for the $\bar\la$. 

Unfortunately it is still not possible to let both terms disappear without making completely unrelated choices for $\Lambda$ and $\bar\Lambda$. As explained above in section \ref{scn:LCcov}, once $\Lambda\neq(\bar\Lambda)^*$, one is dealing with a complexified geometry $SO(4,\ZC)$ and the doubled number of generators of the complexified symmetry group. This is the setup we will be using below.
One convenient choice in order to make $\la^1_2$ and $\bar\la^2_{\dot{2}}$ vanish is the following:
\begin{equation}
\Lambda_\a^{\,\,\b} =\left(
\begin{array}{cc}
1 & 0 \\
-z_1 & 1\\
\end{array}
\right)\quad\text{and}\quad
\bar\Lambda_\ad^{\,\,\bd} =\left(
\begin{array}{cc}
1 & 0 \\
-\bar z_2 & 1\\
\end{array}
\right)
\label{L}\end{equation}
where
\be z_1=\sfrac{\la^1_2}{\la^1_1}\, , \qquad \bar z_2=\sfrac{\bar\la^2_{\dot 2}}{\bar\la^2_{\dot 1}}\,.
\ee
A short calculation reveals that overall momentum is still conserved, all external momenta remain on-shell and the Lorentz-invariant momentum of the internal propagator stays untouched.  In addition, the spinor metric $\epsilon_{\a\b}$ and $\epsilon_{\ad\bd}$ are left invariant. Every particle spinor with the component $2$ or $ \dot 2$ is affected by this transformation, but only  $\la^1_2$ and $\bar\la^2_{\dot{2}}$ vanish after the transformation. This is in contrast to the complex BCFW deformation, which affects only two particular particles. With the transformation (\ref{L}) the $\Delta_{ij}$-terms have disappeared. Still, one is left with the fermionic terms in (\ref{psi}).


\subsubsection{Choice of the fermionic part of a  superframe}
As discussed in chapter 2 of \cite{ArkaniHamed:2008gz}, the set of Grassmann parameters $\eta$ defining a superamplitude is not unique, but exhibits some freedom. In particular, this freedom can be used to set two Grassmann parameters to zero for any $SU(R)$-index $A$
\begin{equation}
\e_{iA}=\e_{jA}=0\quad\forall A\,,
\end{equation}
which can be employed to show the vanishing of the all-plus and all-but-one-plus amplitudes in any supersymmetric theory. Here we will use this freedom to get rid of the unwanted term in the supermomentum conserving $\delta$-function in \eqn{eqn:fourpointresult} by choosing
\be
\eta_{2A}=\eta_{3A}=0\quad\forall A\,,
\ee
which can be achieved by shifting with a supersymmetry parameter
\begin{equation}
\bar \epsilon_{\dot 2}={\eta_2\over \bar \lambda^2_{\dot 1}}\, , \qquad \bar \epsilon_{\dot 1}={\eta_3\over \bar \lambda^3_{\dot 2}}\,.
\end{equation}
Here again the supersymmetric transformation affects all $\eta$'s, but only 2 of the $\eta$'s vanish after the transformation. The overall superframe choice reads 
\begin{equation}\label{superframe}
\boxed {\la^1_2=0 \, , \qquad \bar\la^2_{\dot{2}}=0  \, , \qquad \eta_2=0  \, , \qquad \eta_3=0}\,.
\end{equation}

Having exhausted all freedom to choose a frame, one can now take a look back to the diagrams in the picture above. In order to calculate the color-ordered subamplitude $A(1,2,3,4)$, one does not have to consider diagrams 5 and 6, as those will not contribute to this particular cyclical ordering of legs. This leaves one with diagrams 2,\,3,\,4 and the contact term. Below we will show diagram by diagram that those do not contribute in this particular choice of superframe. 

\paragraph{Diagram 2} This diagram vanishes by a consideration of the supermomentum conserving $\delta$-function at the MHV-vertex.
\begin{equation}
\delta^8(\la^2\eta_2+\la^3\eta_3+\la^P\eta_P)=\delta^8(\la^P\eta_P)=0\,,
\end{equation}
because with only one Grassmann-variable one can not create an eight-dimensional Grassmannian $\delta$-function.
\paragraph{Diagrams 3 and 4}  vanish by consideration of the usual bosonic momentum conservation at the MHV-vertex. The condition
\begin{equation}
\delta(\la^1_2\bar\la^1_{\dot 2}+\la^2_2\bar\la^2_{\dot 2}+\la^P_2\bar\la^P_{\dot 2})=\delta(\la^P_2\bar\la^P_{\dot 2})
\end{equation}
could be satisfied by choosing $\bar\la^P_{\dot 2}=0$, which leads to $\bar\la^1_{\dot 2}=0$ after considering the $1\dot 2$ component of usual momentum conservation at the 12P-vertex. However, this would mean in turn that $[12]=0$ and thus no momentum is transported along the internal propagator. The same is true for the other choice, $\la^P_2=0$, this leads to $\<12\>=0$ (and, in addition $[i2]=0$) by considering now the $2\dot 1$ component of momentum conservation.
\paragraph{Contact terms:}
here the story is a little more intricate. There are six different terms $\psi_{ij}$ in the expression of the contact terms \eqn{eqn:contactterms}. One needs to show that either all of those $\psi$'s disappear or the prefactors should vanish (or a combination of both). With the given kinematical choice, $\la^1_2=0,\,\bar\la^2_{\dot 2}=0$, one can immediately deduce $p^1_+=p^2_+=0$, which lets the two last lines disappear. This can be seen by considering $p^1_+=p^2_+=0$ in the first term of the prefactor in the second and third line: the seemingly complicated expressions yield $1$ or $-1$ respectively, which makes the whole prefactor vanishing because of the Jacobi identity
\begin{equation}
f^{abe}f^{ecd}+f^{bce}f^{ead}+f^{cae}f^{ebd}=0\,.
\end{equation}
Now one is left with the first line, where the term
\begin{equation}
\frac{(p^1-p^2)_+(p^3-p^4)_+}{(p^1+p^2)_+(p^3+p^4)_+}
\end{equation}
blows up. However, one is saved by considering the second factor with the terms $\psi_{12}$ and $\psi_{34}$. While the first one is already zero by our chosen frame, the latter one will read
\begin{equation}
\psi_{34}=\prod_{A=1}^4\eta_{3A}\bar\la^4_{\dot 2}-\eta_{4A}\bar\la^3_{\dot 2}=\prod_{A=1}^4\left(-\eta_{4A}\bar\la^3_{\dot 2}\right)\,.
\end{equation}
Considering now the supermomentum conservation, one finds
\begin{equation}
\delta^4(\la^1_2\eta_1+\la^2_2\eta_2+\la^3_2\eta_3+\la^4_2\eta_4)=\delta^4(\la^4_2\eta_4).
\end{equation}
While this expression alone could be satisfied, one can rewrite any Grassmanian $\delta$ by its argument and vice versa. The expression
\begin{equation}
\delta^4(\eta_4\bar\la^3_{\dot 2})\delta^4(\la^4_2\eta_4),
\end{equation}
does not represent a possible contribution, as again with only four Grassmann variables $\eta_4$ an eight-dimensional Grassmannian $\delta$-function shall be saturated. Thus the expected result is from the first diagram is really the only contribution.

Note however, that we have used here that we already knew the result to be Lorentz invariant. Otherwise the algebra would have been more involved.


\subsection{\Nf SYM $n$-point MHV amplitudes in a preferred superframe}

While for the four-point amplitude the choice of a superframe and the corresponding simplifications have been described in detail, this subsection is devoted to the generalization of this scenario to higher-point MHV amplitudes. We will show how all MHV-amplitudes can be generated recursively by adding a $\overline{\text{MHV}}$-vertex to an $(n-1)$-point MHV amplitude yielding a $n$-point MHV amplitude. 

In order to reuse the same arguments as in the previous subsection, we will choose the $(n-1)$-point MHV amplitude to be labelled by $(1, P, 4,...,n)$ while the three-point ${\overline{\text{MHV}}}$ vertex has labels $(2,3,P)$. Accordingly, the $(n-1)$-point MHV amplitude comes with the $\delta$-function of the form $\delta^4(p_1+p_n+... +P) \delta^8 (\lambda^1\eta_1+ \lambda^n\eta_n+ ... + \lambda^P \eta_P )$ and the three-point ${\overline{\text{MHV}}}$ vertex $(2,3,P)$ contributes a $\delta$-function of the form
$\delta^4(p_2+p_3-P) \delta^4( \eta_P [23] + \eta_2[3P] +\eta_3 [P2])$. 

In the same way as was done in \cite{Fu:2010qi}, one can now join the two fermionic $\delta$-functions from both vertices. Again, there will be additional terms:
\begin{equation}\label{eqn:npointresult}
\delta^4 \left (\sum _{i=1}^n \la^i_2 \eta_i  \right) \delta^4\left (\sum _{i=1}^n \la^i_1 \eta_i -{\eta_2 ( \xi^P \bar \xi^P \bar \la^3)_1  - \eta_3 ( \xi^P \bar \xi^P \bar \la^2)_1 \over [23]} \right),
\end{equation}
while the bosonic momentum conservation will read $\delta^4( \sum _{i=1}^n p_i)$. So the numerator of the $n$-point generalization of \eqn{eqn:fourpointresult} remains unchanged up to altered summation limits. As in the four-point-case, the $\eta$-integration delivers a factor of $[23]^4$.
The remaining factor to evaluate now reads:
\begin{equation}
\frac{1}{\<45\> \<56\> ...\<n-1, n\> \<n1\>}\frac{[23]^4}{P^2 \<41\>[23] \< 1P\> \< P4\> [P2] [3P]}.
\end{equation}
Replacing the propagator by $\<23\>[23]$ and employing momentum conservation, for example
\begin{eqnarray}
\<12\>[23]+\<14\>[43]+\cdots +\<1n\>[n3]&=&\<12\>[23]-\<1P\>[P3]-\<1\xi\>[\xi 3]\nnl&=&\<12\>[23]-\<1P\>[P3]-\la^1_2\xi_1\bar\xi_{\dot 1}\bar\la^3_{\dot 2}\nnl&=&0
\end{eqnarray}
yields the expression 
\begin{equation}
\frac{1}{\<23\>\<45\>\cdots\<n\,1\>(\<12\>-\Delta_{13})( \<34\>-\Delta_{42})}\,.
\end{equation}
Thus the additional angular brackets in the denominator and in the fermionic $\delta$-functions are spectators - they do not influence the Lorentz non-covariant parts. 

Noting this, it is clear that the same arguments as for the four-point function can be applied. With exactly the same choice of a superframe the Lorentz non-covariant terms can be made disappear thus producing the correct expression for the $n$-point superamplitude
\begin{equation}
\frac{\delta^4\left(\sum_{i=1}p_i\right)\delta^4 \left (\sum _{i=1}^4 \la^i_2 \eta_i  \right) \delta^4\left (\sum _{i=1}^4 \la^i_1 \eta_i \right)}{\<12\>\<23\>\<34\>\<45\>\cdots\<n\,1\>}\,.
\end{equation}

This in turn means that the superframe choice has sufficiently ruled out all other possible contributions to the five-point amplitude. However, this is difficult to show explicitely. 
Generalizations to more legs are straightforward. Starting from the four-point function one can obtain all MHV superamplitudes in \Nf SYM recursively. In the light of the similarities to the BCFW formalism discussed above this is not surprising, but here one can reach the same goal with the choice of a superframe. An important point to note is the following: a necessary condition for the valid choice of a superframe (or a two-line BCFW-shift) is the existence of just \textit{one} internal propagator. 

Whether the method works for the NMHV sector remains to be shown. If we were to speculate, a modest guess would be that the method works in any situation with just one internal propagator, thus mimicking the BCFW-situation.


\section{Amplitudes in \Ne}

\subsection{\Ne supergravity 4-point amplitude in a preferred superframe}

Noting the three-point interactions in \Ne supergravity being the squares of the corresponding expressions in \Nf SYM, the expression to calculate for the first supergraph in (\ref{eqn:fourpointpicture}) reads (cf. \eqn{BHT1})
\be
{\delta^4(p_1+p_4+P) \delta^{16} (\lambda^1\eta_1+ \lambda^4 \eta_4 + \lambda^P \eta_P ) \delta^4(-P+p_2+p_3)  \delta^8( \eta_P [23] + \eta_2[3P] +\eta_3 [P2])\over P^2 \left\langle 41\right\rangle^2 [23]^2 \left\langle 1P\right\rangle^2 \left\langle P4\right\rangle^2 [P2]^2 [3P] ^2}
\label{BHT2}\ee
which has to be integrated over $d^4Pd^8\eta$. Again, bosonic momentum conservation is obvious and the fermionic momentum conservation will work the same way as in the \Nf SYM case. From the fermionic integration however, one will now have a factor of $[23]^8$. In order to obtain exactly the square of \eqn{eqn:fourpointresult} the propagator would have to be squared as well. This not being the case the integration over \eqn{BHT2} will produce
\be
M(1,2,3,4) =t A^2 (1,2,3,4) = t\left(\frac{1}{\<12\>\<23\>\<34\>\<41\>}\right)^2= {[34]^4\over \left\langle 12\right\rangle^4} {t\over s^2t^2}
\label{1234}\ee
where the factor $t=s_{23}$ compensates for the missing squaring of the propagator.
In the above calculation producing $M(1,2,3,4)$ we have been using the same choice of a superframe \eqn{superframe} as in the previous section, which lets the additional terms disappear.  

Obviously, the above expression is not totally symmetric under all permutations of the legs $1,2,3$ and $4$. In order to account for the complete amplitude, 
\be
W^4(\phi_{\rm in})=   \int \prod_{i=1}^4 \{d^8 z_i \phi_{\rm in}(z_i)\} (2\pi)^4 \delta^4(\sum p_i) \delta^{16}(\sum_i\lambda^i \eta_i) {\cal M }(1,2,3,4)
\label{W4}\ee
one has to symmetrize $M (1,2,3,4)$ to obtain the totally symmetric amplitude {\cal M }(1,2,3,4). One way to do this is to recognize that there is a second diagram with propagator $s$ (diagram number 3 in figure in (\ref{eqn:fourpointpicture}), which contributes to the amplitude $A(1,2,3,4)$. So the object which needs to be symmetrized over all available distinct orderings is $ {[34]^4\over \< 12\>^4} {t+s\over s^2t^2}$. Thus
\begin{eqnarray}\label{M1234}
{\cal M }(1,2,3,4)&=&\frac{1}{3}  {[34]^4\over \left\langle 12\right\rangle^4} \left[{t+s\over s^2t^2}+ {s+u \over s^2u^2}+ {t+u\over u^2t^2}\right]\nnl&=&-\frac{1}{3}
{[34]^4\over \left\langle 12\right\rangle^4} {u^3 +s^3 + t^3 \over s^2t^2 u^2} \nnl&=&-{1\over stu} {[34]^4\over \left\langle 12\right\rangle^4}
\end{eqnarray}
Of course one could have also calculated all three different color-orderings and then averaged over those in order to obtain the same result. 

\subsection{Contact terms in \Ne supergravity}

The calculation of amplitudes in \Ne supergravity for higher multiplicities using the Feynman rules originating from the new action \eqn{eqn:N8action} will be left for future work. At tree level, these amplitudes were studied in \cite{Elvang:2007sg,Bianchi:2008pu,Drummond:2009ge} using information about their symmetries, different forms of KLT-like relations and recursion relations. One of the reasons besides the algebraic complexity which prevents the use of a path integral approach is the fact that contact terms in \Ne supergravity are not known explicitely. 

In section \ref{scn:amplitudes} above it has been shown that the choice of a preferred superframe is capable of removing contact term contributions in \Nf SYM theory for MHV amplitudes. The same is true in a broader situation for the momentum deformation in the BCFW scenario, in which the answer for all tree amplitudes in \Nf SYM and \Ne supergravity can be obtained recursively without reference to anything but the three-point interactions.  

Naturally, contact terms can not be neglected in the calculation of Feynamn diagrams. On the other hand, their contributions can be made disappear in any step of a recursive calculation. This in turn means, that no information on contact terms in the actions for \Nf SYM and \Ne supergravity is required, only the three-point vertices in the actions \eqns{eqn:N4action}{eqn:N8action} are relevant given that the proper procedure of calculation is established.

So contact term information is not required in order to obtain an answer for all tree amplitudes in \Ne supergravity and \Nf SYM. If one, however, wants to proceed to loops, one can for example employ unitarity techniques, which will be discussed in the next section in the context of the double-copy property.


\section{\Ne supergravity as a double copy of \Nf SYM}\label{scn:double}

The actions in \eqns{eqn:N4action}{eqn:N8action} for \Nf SYM and \Ne supergravity derived in the last section do not posess any gauge symmetries any more, thus the path integral is just the exponent of those actions accompanied by a source term. Considering the form of the three-point vertices in the actions, one is immediately reminded of the double-copy property relating \Nf SYM theory and \Ne supergravity. Before commenting on possible connections, we want to state the basic setup for the underlying color-kinematic duality. Although having originated in a setup with gauge theory and gravity amplitudes initially \cite{Bern:2008qj}, the techniques have been shown to be applicable and valid for relating two copies of \Nf SYM theory to \Ne supergravity \cite{Bern:2010ue,Bern:2010fy,Bern:2010yg}.

An arbitrary superamplitude in \Nf SYM theory is obtained by summing over all contributing topologies allowed by the corresponding Feynman rules. After absorbing any higher vertex into three-vertices with the help of additional fields, one can express the superamplitude as

\begin{equation}
(-i)^L {\cal A} ^{\rm loop}_{n}= \sum_j \int \prod_{l=1}^L {d^4p_l\,d^4\e_l \over (2\pi)^4}  {1\over s_j} {n_j c_j\over \prod_{\alpha_j} p^2_{\alpha_j}},
\label{BCJ1}
\end{equation}
where the summation over $j$ runs over all contributing cubic diagrams and $\prod_{\alpha_j} p^2_{\alpha_j}$ is a product of all propagators of a particular diagram $j$. Furthermore, $c_j$ is the color factor of the entire diagram obtained by calculating the particular combination of structure constants for a diagram $j$ and $n_j$ is called the numerator factor collecting the dependencies on the external momenta $p$. Finally, $s_j$ is the symmetry factor associated with the particular diagram $j$. 

The numerator factor $n_j$ for a particular cubic diagram $j$ can be determined as follows: allowing for complex momenta, one can always find a kinematical configuration\footnote{Performing the calculation in dimensions $D>4$ might be required.}, in which all internal propagators are on-shell. This configuration is referred to as the maximal unitary cut and provides the first contribution to the numerator factor $n_j$:
\begin{equation}\label{eqn:defnj}
n_j=(\text{max. cut})+\sum_{P\in p_{\a_j}}P^2\cdot(\text{next-to-max. cut})+\cdots\,.
\end{equation}
Since in the maximal cut scenario all propagators are on-shell, one misses terms proportional to $P^2$. Those can be obtained by releasing cut conditions, which means leaving certain propagators off-shell. If just one propagator remains off-shell, this is referred to as next-to-maximal cut and so on. 
In the maximal cut situation only three point interactions are involved and thus the evaluation of the numerator factor in \eqn{BCJ1} boils down to multiplying three vertices with deformed momenta satisfying the maximal cut conditions. For the next-to-maximal cuts the situation is not that straightforward: here higher-multiplicity subamplitudes can occur, for which all contributing diagrams have to be considered. \Eqn{eqn:defnj} is not an unambiguous definition of the numerator factor $n_j$. As there is some residual gauge freedom in the $n_j$, one can assign the contributions from non-maximal cuts to different $n_j's$. 

One of the statements of \cite{Bern:2010fy,Bern:2010yg} is that one can find a set of numerator factors $n_j$ in \Nf SYM theory such that amplitudes in \Ne supergravity can be expressed by \eqn{BCJ1} with the color factor $c_j$ for each diagram being replaced by a second numerator factor $\tilde n_j$:
\begin{equation}
(-i)^{L+1}  {\cal M}^{\rm loop}_{n}= \sum_j \int \prod_{l=1}^L {d^4p_l\,d^8\e_l\over (2\pi)^4}  {1\over S_j} {n_j \tilde n_j\over \prod_{\alpha_j} p^2_{\alpha_j}}\,.
\label{BCJ2}
\end{equation}
Here the additional factor of $i$ compared to \eqn{BCJ1} is conventional and  the second set $\tilde n_j$ can be chosen to be equal $n_j$. 

However, this double-copy property does not hold for any set of numerator factors $n_j$ in \Nf SYM theory. In order for \eqn{BCJ2} to yield the correct \Ne supergravity amplitude, the $n_j$'s have to satisfy relations similar to the Jacobi identities for the color-factors. This means that any set of three numerators $n_j$ whose corresponding diagrams differ only by the form of a particular four-point subdiagram have to satisfy the same relation as the associated color-factors do \cite{Bern:2008qj}. Working with a set $n_j$ satisfying the Jacobi identities is a sufficient condition for \eqn{BCJ2} to be valid. It demands the non-maximal contributions to be distributed onto different $n_j$'s in a particular way. An example of different choices for the set of $n$'s can be found comparing the two papers \cite{Bern:2007hh} and \cite{Bern:2010ue}. In the first version, the calculated numerator factors in \Ne supergravity have not been the square of the corresponding numerators in \Nf SYM off-shell for all diagrams. This was due to the fact that the non-maximal cut contributions had not been distributed onto the numerator factors of \Nf SYM in the way found in \cite{Bern:2010ue} later. 

How is the double-copy property related to the new actions? While the kinetic terms of both theories coincide (up to doubling the number of Grassmann variables), the three-point interactions in \eqns{eqn:N4action}{eqn:N8action}, mimic
the behaviour of the two general expressions \eqns{BCJ1}{BCJ2} with additional Grassmann $\eta$-integration.
\begin{center}
\begin{tabular}{cc|cc}
\Nf&\hspace{.1cm}&\hspace{.1cm}&\Ne\\\hline\hline&&&\\[-12pt]
${n_j c_j\over \prod_{\alpha_j} p^2_{\alpha_j}}$&&&${n_j \tilde n_j\over \prod_{\alpha_j} p^2_{\alpha_j}}$\\[2pt]\hline&&&\\[-12pt]
$\frac{f^{abc}\delta^8(\sum_i\la^i\eta_i)}{\<12\>\<23\>\<31\>}+\frac{f^{abc}\delta^4(\sfrac{1}{2}\ve^{ijk}[ij]\eta_k)}{[12][23][31]}$&&&$\frac{\delta^{16}(\sum_i\la^i\eta_i)}{(\<12\>\<23\>\<31\>)^2}+\frac{\delta^8(\sfrac{1}{2}\ve^{ijk}[ij]\eta_k)}{([12][23][31])^2}$
\end{tabular}
\end{center}
Comparing the third row in the above table, it is clear that in order to obtain the three-point interactions of \Ne supergravity one has to remove the color part of the \Nf SYM vertex and square the remainder. In addition, the $\delta$-functions ensuring supermomentum conservation have to be promoted to accomodate twice the number of Grassmann variables. Crossterms do not occur as those would have the wrong dimension of supermomentum conserving $\d$-functions. As any of the vertices is contracted with the appropriate lightcone superfields, the vertices in either theory should be completely symmetric. While in \Nf SYM theory this is obtained by multiplying two different completely antisymmetric objects, $f^{abc}$ and $\sfrac{1}{\<ij\>\<jk\>\<ki\>}$ or $\sfrac{1}{[ij][jk][ki]}$, and tying their indices pairwise to the Lie-algebra valued superfield $\phi^{ia}$, the expressions $\sfrac{1}{(\<ij\>\<jk\>\<ki\>)^2}$ and $\sfrac{1}{([ij][jk][ki])^2}$ are totally symmetric by construction.

Looking back to \eqn{eqn:defnj} and comparing the situations for the two maximally supersymmetric theories, one immediately finds that the maximal cut contribution to $n_j$ for \Ne supergravity can be obtained from the maximal cut in \Nf SYM theory by stripping off the color factor $c_j$ and squaring $n_j$, since all internal propagators are on-shell. 

For the non-maximal cut contributions, however, this is not necessarily true: individual graphs in \Ne supergravity are not the square of the corresponding \Nf graph in general. Being proportional to $P^2$ (or higher powers of internal momenta), the non-maximal cut contributions have the same form as the contact terms. A set of \Nf SYM numerators satisfying the the same Jacobi identities as their associated color factors is not unique, but still exhibits some remaining freedom. It is this generalized gauge freedom, which can be used to reshuffle the contact contributions between different $n_i$'s while maintaining the double-copy property. 

In the new path integral approach, supersymmetry breaking terms appear, which have been shown to exhibit the same form as contact term (see subsection \ref{ssn:susbreak} above).  These terms have to disappear in the final result to yield a supersymmetric answer, which means that supersymmetry-breaking terms and contact terms have to cancel after considering all contributing diagrams. Although being similar in structure to the $P^2$-proportional next-to-maximal cut contributions, they can not be shuffled around here: they are unambigously assigned to certain diagrams in \Nf SYM theory by the Feynman rules derived from the action \eqn{eqn:N4action}. It would be nice to work out, whether this particular distribution of contact-type terms is related to the Jacobi identities for the numerators in any way. 


\section{Discussion}
In this article, actions for \Nf SYM and \Ne supergravity have been formulated in lightcone superspace in a way favourable for on-shell amplitudes and recursion relations. The vertices are of the known form of the MHV and $\overline{\text{MHV}}$ three-point interactions in either theory. In order to be part of a path integral formulation, they should also hold as off-shell expressions. Therefore the notion of helicity brackets had to be extended to off-shell vectors promoting the vertices to projections onto the on-shell parts of the contributing legs. 

The calculation of amplitudes in the new framework is closely related to recursion relations. Choosing a preferred superframe in \Nf SYM one can simplify the calculation of MHV diagrams considerably. Although one is reminded of BCFW recursion relations, the procedure employed here is different in the sense that a superframe is chosen in contrast to the complex deformation in the BCFW case. However -- as in the BCFW scenario -- this particular superframe choice excludes certain diagrams from the calculation. And, quite similarly, our vertices are on-shell projections of the off-shell vertices. So our formalism assumes that one can bring all internal propagators on-shell. In addition, one has to note that the choice of a preferred superframe has only been successful in the case of just one internal propagator. If two internal propagators are present, the choice is not sufficient. This reminds once again on the scenario in the BCFW recursion relation and on more general recursions.

The actions found are suggestive for the double-copy relation of \Nf SYM and \Ne supergravity, as the supergravity three-vertices are the square of the \Nf three-vertices after stripping of the color information. While for maximal cuts in purely cubic diagrams the double-copy property is obvious, non-maximal cuts have to be organized in a way that the numerator factors for \Nf SYM satisfy the appropriate Jacobi identities for the double-copy property to hold. Noting that non-maximal cuts are algebraically on the same footing as contact term contributions and supersymmetry-violating terms in the lightcone three-vertices, it would be interesting to find out how the requirement of unbroken supersymmetry in the final amplitude is possibly related to the Jacobi identities for the numerators.


\section*{Acknowledgments}
We are grateful to J.J Carrasco, H. Johansson, J. Kaplan and M. Kiermaier for stimulating discussions. This work is supported by the NSF grant 0756174. J.B. acknowledges support from the Alexander-von-Humboldt foundation within the Feodor-Lynen program. 

\appendix
\section{Relating different versions of the \Nf SYM lightcone path integral}


\subsection{Three-point $\overline{\text{MHV}}$ vertex}
Leaving the prefactor of $-2ig(2\pi)^4f^{abc}$ aside for a moment, one starts with the first part of \eqn{eqn:BDKMvertices} accompanied with the appropriate $\delta$-functions
\begin{equation}
\{12\}\delta^{(4)}\left(\sum\limits_{i=1}^3 \pi_{i}\right)\delta^{(4)}\left(\sum\limits_{i=1}^3 p_{i}\right)=\{12\}\delta^{(4)}\left(\sum\limits_{i=1}^3 \eta_{i}\sqrt{p_{i+}}\right)\delta^{(4)}\left(\sum\limits_{i=1}^3 p_{i}\right)\,.
\end{equation}
Pulling out a factor of $\sqrt{(p_{1+}p_{2+}p_{3+})}$ from the argument of the supermomentum conserving delta function leads to
\begin{equation}
\{12\}(p_{1+}p_{2+}p_{3+})^2\delta^{(4)}\left(
\frac{\eta_1}{\sqrt{p_{2+}p_{3+}}}+
\frac{\eta_2}{\sqrt{p_{3+}p_{1+}}}+
\frac{\eta_3}{\sqrt{p_{1+}p_{2+}}}\right)\delta^{(4)}\left(\sum\limits_{i=1}^3p_{i}\right)\,.
\end{equation}
Employing definition \eqn{eqn:spinorbrackets}, any of the factors $1/\sqrt{p_{i+}p_{j+}}$ can be replaced by the ratio $[ij]/\sqrt{2}\{ij\}$, which leads to:
\begin{equation}
\frac{\{12\}}{(\sqrt{2})^4}(p_{1+}p_{2+}p_{3+})^2\delta^{(4)}\left(
\eta_1\frac{[23]}{\{23\}}+
\eta_2\frac{[31]}{\{31\}}+
\eta_3\frac{[12]}{\{12\}}\right)\delta^{(4)}\left(\sum\limits_{i=1}^3p_{i}\right)
\end{equation}
By momentum conservation \eqn{eqn:momentumconservation} one sees $\{12\}=\{23\}=\{31\}$, which allows pulling out the denominators from the argument of the delta function:
\begin{equation}
\frac{\{12\}}{(\sqrt{2})^4\,\{12\}^4}(p_{1+}p_{2+}p_{3+})^2\delta^{(4)}\left(
\eta_1[23]+
\eta_2[31]+
\eta_3[12]\right)\delta^{(4)}\left(\sum\limits_{i=1}^3p_{i}\right)\,.
\end{equation}
Cancelling one factor of $\{12\}$, using again $\{12\}^3=\{12\}\{23\}\{31\}$, replacing the curly brackets by the angular ones with the help of \eqn{eqn:spinorbrackets} and restoring the prefactor finally leads to
\begin{equation}
g\sqrt{2}(2\pi)^4f^{abc}\frac{(p_{1+}p_{2+}p_{3+})}{i}\frac{\delta^{(4)}\left(
\eta_1[23]+
\eta_2[31]+
\eta_3[12]\right)\delta^{(4)}\left(\sum_{i=1}^3p_{i}\right)}{[12][23][31]}.
\end{equation}
Taking into account the different definition of fields in BDKM and Fu/Kallosh one has to divide by
\begin{equation}
\left(\frac{-i}{p_{1+}}\right)\left(\frac{-i}{p_{2+}}\right)\left(\frac{-i}{p_{3+}}\right)\cdot\frac{i}{p_{1+}p_{2+}p_{3+}}
\end{equation}
which finally leads to
\begin{equation}
g\sqrt{2}(2\pi)^4f^{abc}\frac{\delta^{(4)}\left(
\eta_1[23]+
\eta_2[31]+
\eta_3[12]\right)\delta^{(4)}\left(\sum_{i=1}^3p_{i}\right)}{[12][23][31]},
\end{equation}
where the different prefactor in comparison with \eqn{eqn:N4action} originates in the different normalization of the structure constants $f^{abc}$ and a symmetry factor which is not included in the vertex \eqn{eqn:BDKMvertices}.

\subsection{Three-point $\text{MHV}$ vertex}

Starting now from
\begin{equation}
-2ig(2\pi)^4f^{abc}\frac{(12)\ki{12}}{(p_{1+}p_{2+}p_{3+})^2}\,\delta^{(4)}\left(\sum\limits_{i=1}^3 \pi_{i}\right)\delta^{(4)}\left(\sum\limits_{i=1}^3 p_{i}\right)\,,
\end{equation}
one can first transform the fermionic bracket by means of \eqn{eqn:fermbracketsresolve}
which leads to
\begin{equation}
-2ig(2\pi)^4f^{abc}\frac{(12)}{p_{3+}^2}\,\prod\limits_{A=1}^{4}(\eta_{1,A}\sqrt{p_{2+}}-\eta_{2,A}\sqrt{p_{1+}})
\delta^{(4)}\left(\sum\limits_{i=1}^3 \eta_{i}\sqrt{p_{i+}}\right)\delta^{(4)}\left(\sum\limits_{i=1}^3 p_{i}\right)\,,
\end{equation}
where again the argument of the supermomentum-conserving delta funtion was transformed employing \eqn{eqn:fermrelation}. Considering now the fermionic part exclusively one finds
\begin{eqnarray}
&&\prod\limits_{A=1}^{4}(\eta_{1,A}\sqrt{p_{2+}}-\eta_{2,A}\sqrt{p_{1+}})
\delta^{(4)}\left(\sum\limits_{i=1}^3 \eta_{i}\sqrt{p_{i+}}\right)\nnl
&=&\prod\limits_{A=1}^{4}(\eta_{1,A}\sqrt{p_{2+}}-\eta_{2,A}\sqrt{p_{1+}})
\prod\limits_{A=1}^{4}\left(\sum\limits_{i=1}^3 \eta_{i,A}\sqrt{p_{i+}}\right)\nnl
&=&\prod\limits_{A=1}^{4}\left(
-\eta_{1,A}\eta_{2,A}(p_{1+}+p_{2+})
+\eta_{3,A}\eta_{1,A}\sqrt{p_{3+}}\sqrt{p_{2+}}
-\eta_{3,A}\eta_{2,A}\sqrt{p_{3+}}\sqrt{p_{1+}}
\right)\nnl
&=&\prod\limits_{A=1}^{4}\left(
\eta_{1,A}\eta_{2,A}\sqrt{p_{3+}}\sqrt{p_{3+}}
+\eta_{3,A}\eta_{1,A}\sqrt{p_{3+}}\sqrt{p_{2+}}
-\eta_{3,A}\eta_{2,A}\sqrt{p_{3+}}\sqrt{p_{1+}}
\right)\nnl
&=&(p_{1+}p_{2+}p_{3+})^2\,p_{3+}^2\,\prod\limits_{A=1}^{4}\left(
\frac{\eta_{1,A}\eta_{2,A}}{\sqrt{p_{1+}p_{2+}}}
+\frac{\eta_{2,A}\eta_{3,A}}{\sqrt{p_{2+}p_{3+}}}
+\frac{\eta_{3,A}\eta_{1,A}}{\sqrt{p_{3+}p_{1+}}}
\right).
\end{eqnarray}
Now one can play the same game as for the $\overline{\text{MHV}}$-vertex by replacing
\begin{equation}
\frac{1}{\sqrt{p_{i+}p_{j+}}}=\frac{\<ij\>}{\sqrt{2}(ij)}\,,
\end{equation}
which after pulling out factors of $(12)=(23)=(31)$ yields
\begin{equation}
\frac{(p_{1+}p_{2+}p_{3+})^2\,p_{3+}^2}{(\sqrt{2})^4(12)^4}\,\prod\limits_{A=1}^{4}\left(
\eta_{1,A}\eta_{2,A}\<12\>
+\eta_{2,A}\eta_{3,A}\<23\>
+\eta_{3,A}\eta_{1,A}\<31\>
\right)\,,
\end{equation}
where the product can be rewritten as
\begin{equation}
\prod\limits_{A=1}^{4}\left(
\eta_{1,A}\eta_{2,A}\<12\>
+\eta_{2,A}\eta_{3,A}\<23\>
+\eta_{3,A}\eta_{1,A}\<31\>
\right)=\prod\limits_{A=1}^{4}\sum_{i,j=1}^3\,\eta_{i,A}\eta_{j,A}\<ij\>=\delta^{(8)}(\sum_{i+1}^3\la_i\eta_i)\,.
\end{equation}
Putting now everything together will yield
\begin{equation}
-2ig(2\pi)^4f^{abc}\frac{(12)}{p_{3+}^2}\,
\frac{(p_{1+}p_{2+}p_{3+})^2\,p_{3+}^2}{(\sqrt{2})^4(12)^4}\,
\delta^{(8)}\left(\sum_{i=1}^3\la_i\eta_i\right)
\delta^{(4)}\left(\sum\limits_{i=1}^3 p_{i}\right)\,,
\end{equation}
which is the same as
\begin{equation}
g\sqrt{2}(2\pi)^4f^{abc}\frac{(p_{1+}p_{2+}p_{3+})}{i}
\,\frac{\delta^{(8)}\left(\sum_{i=1}^3\la_i\eta_i\right)
\delta^{(4)}\left(\sum_{i=1}^3 p_{i}\right)}{\<12\>\<23\>\<31\>}.
\end{equation}
Taking again the different definitions of the lightcone superfield in \cite{Fu:2010qi} and \cite{Belitsky:2004sc} into account, one is left with
\begin{equation}
g\sqrt{2}(2\pi)^4f^{abc}
\,\frac{\delta^{(8)}\left(\sum_{i=1}^3\la_i\eta_i\right)
\delta^{(4)}\left(\sum_{i=1}^3 p_{i}\right)}{\<12\>\<23\>\<31\>},
\end{equation}
which agrees up to a numerical prefactor (see the last remark in the previous subsection) with \eqn{eqn:N4action}.

\end{document}